\documentclass[sigconf, nonacm]{acmart}
    
\newcommand\vldbdoi{XX.XX/XXX.XX}

\newcommand\vldbavailabilityurl{https://github.com/Yuyang-Song/QUITE}
\newcommand\vldbpagestyle{plain} 

\usepackage{microtype}
\newcommand{\advice}{\mathcal{A}}

\usepackage[linesnumbered,ruled,vlined]{algorithm2e} 
\usepackage{epstopdf}
\usepackage{multicol}
\usepackage{url}
\usepackage{enumerate}
\usepackage[labelfont=bf]{caption}
\usepackage{xcolor}
\usepackage{color, colortbl}
\usepackage{hyperref}
\usepackage{algpseudocode}
\usepackage{tabularx}
\usepackage{makecell} 
\usepackage{subfig}
\usepackage{graphicx}
\usepackage{pifont}
\usepackage{enumitem} 

\usepackage{multirow}
\usepackage{diagbox}

\usepackage{boldline}
\definecolor{darkgreen}{HTML}{006400}

\newcommand{\cmark}{\textcolor{darkgreen}{\ding{51}}}%
\newcommand{\xmark}{\textcolor{red}{\ding{55}}}%

\usepackage{soul} 
\newtheorem{example}{\textbf{Example}}
\AtBeginDocument{%
  }

\usepackage{listings} 
\usepackage{xcolor} 
\usepackage{amsthm}           
\usepackage{xcolor}
\usepackage[most]{tcolorbox}
\usepackage{fontawesome5}
\usepackage{algpseudocode}
\usepackage{makecell}

\definecolor{boxbg}{RGB}{248,250,252} 
\definecolor{boxframe}{RGB}{68,110,155} 
\definecolor{titlecolor}{RGB}{51,101,138} 

\newtcolorbox{promptbox}[2][]{%
    enhanced,
    unbreakable,
    before skip=2mm,
    after skip=2mm,
    colback=boxbg,
    colframe=boxframe,
    coltitle=white,
    boxrule=0.4mm,
    sharp corners,
    arc=2pt,
    attach boxed title to top center={yshift=-2mm},
    boxed title style={
        enhanced,
        colback=titlecolor,
        colframe=titlecolor,
        arc=2pt,
        outer arc=2pt,
        boxrule=0pt,
    },
    title={\faLightbulb[solid]\space #2},
    fonttitle=\bfseries\color{white},
    #1
}

\theoremstyle{definition}     

\newcommand{\system}{{\textsc{QUITE}}}

\begin{document}
\title{QUITE: A Query Rewrite System Beyond Rules via LLM Agents}

\author{Yuyang Song}
\orcid{0009-0009-6571-9036}
\affiliation{%
  \institution{Sichuan University}
  \city{Chengdu, Sichuan}
  \country{China}
}
\email{yuyangsong004@gmail.com}

\author{Hanxu Yan}
\orcid{0009-0000-0140-6350}
\affiliation{%
  \institution{Sichuan University}
  \city{Chengdu, Sichuan}
  \country{China}
}
\email{hanxuyan888@gmail.com}

\author{Jiale Lao}
\orcid{0009-0003-1144-5152}
\affiliation{%
  \institution{Cornell University}
  \city{Ithaca, New York}
  \country{USA}
}
\email{jiale@cs.cornell.edu}

\author{Yibo Wang}
\orcid{0009-0005-1971-3398}
\affiliation{%
  \institution{Purdue University}
  \city{West Lafayette, Indiana}
  \country{USA}
}
\email{wang7342@purdue.edu}

\author{Yufei Li}
\orcid{0009-0004-4285-5696}
\affiliation{%
  \institution{Hong Kong University of Science and Technology}
  \city{Hong Kong SAR}
  \country{China}
}
\email{liyufeievangeline@gmail.com}

\author{Yuanchun Zhou}
\orcid{0000-0003-2144-1131}
\affiliation{%
  \institution{Computer Network Information Center, Chinese Academy of Sciences}
  \city{Beijing}
  \country{China}
}
\email{zyc@cnic.cn}

\author{Jianguo Wang}
\orcid{0000-0002-3039-1175}
\affiliation{%
  \institution{Purdue University}
  \city{West Lafayette, Indiana}
  \country{USA}
}
\email{csjgwang@purdue.edu}

\author{Mingjie Tang}
\orcid{0000-0002-8893-4574}
\affiliation{%
  \institution{Sichuan University}
  \city{Chengdu, Sichuan}
  \country{China}
}
\email{tangrock@gmail.com}

\begin{abstract}
Query rewrite transforms a given SQL query into a semantically equivalent query that can be executed more efficiently. Existing approaches mainly rely on predefined rewrite rules. However, they can only handle a small subset of queries and may lead to performance regressions. 
This limitation arises from the intrinsic challenges of rule-based query rewrite: (1) it is hard to discover and verify new rewrite rules, (2) fixed rewrite rules do not generalize to new query patterns and are insufficient for handling complex queries, and (3) some rewrite techniques cannot be expressed by fixed rules. 
Motivated by the fact that human experts exhibit significantly better rewrite ability but suffer from scalability, and Large Language Models (LLMs) have demonstrated nearly human-level semantic and reasoning abilities, we propose a new approach of using LLMs to rewrite SQL queries beyond rules. Due to the hallucination problems in LLMs, directly applying LLMs often leads to nonequivalent and suboptimal queries. To address this issue, we propose \system\ (\underline{qu}ery rewr\underline{ite}), a training-free and feedback-aware system based on LLM agents that rewrites SQL queries into semantically equivalent forms with significantly better performance, covering a broader range of query patterns and rewrite strategies compared to rule-based methods. 
Firstly, we design a multi-agent framework controlled by a finite state machine (FSM) to equip LLMs with the ability to use external tools and enhance the rewrite process with real-time database feedback. 
Secondly, we develop a rewrite middleware to enhance the ability of LLMs to generate optimized query equivalents. It includes a structured knowledge base for domain knowledge, an SQL corrector to ensure equivalence, and an agent memory buffer to manage critical rewrite contexts. 
Finally, we employ a novel hint injection technique to produce better execution plans for rewritten queries. Extensive experiments show that our method achieves up to a \textbf{35.8\%} reduction in query execution over state-of-the-art approaches. Moreover, it delivers \textbf{24.1\%} additional rewrites compared to prior methods, which cover query cases that earlier approaches could not handle.
\end{abstract}

\maketitle

\pagestyle{\vldbpagestyle}



\vspace{-0.5em}
\section{Introduction}\label{sec: intro}
Inefficient SQL queries remain a long-standing challenge in many database applications, often arising from poorly written queries by inexperienced users or automatically generated systems~\cite{cloud,database,extensible}.
Query rewrite is a promising approach that transforms an original SQL into a semantically equivalent form with better performance~\cite{query-rewrite,extensible,wetune}. An effective rewritten query can dramatically improve the execution time of a poorly written SQL query by several orders of magnitude~\cite{wetune,lr}.

Existing query rewrite approaches typically rely on predefined rewrite rules~\cite{apache,volcano,pg,lr,lr_demo, querybooster,llm-r2,r-bot}. Heuristic-based methods apply these rules in a fixed order derived from practical experience~\cite{apache,volcano,pg}. However, such fixed orders often fail to generalize across different environments and query patterns. Thus, machine learning (ML)-based methods train models on rewrite histories to predict which rules to use and the order in which to apply them~\cite{lr,lr_demo}. Recently, some approaches have explored the use of Large Language Models (LLMs) to select rewrite rules and determine their application order based on domain knowledge~\cite{llm-r2,r-bot}. 

\begin{figure*}[t]
    \centering
    \includegraphics[width=0.95\linewidth]{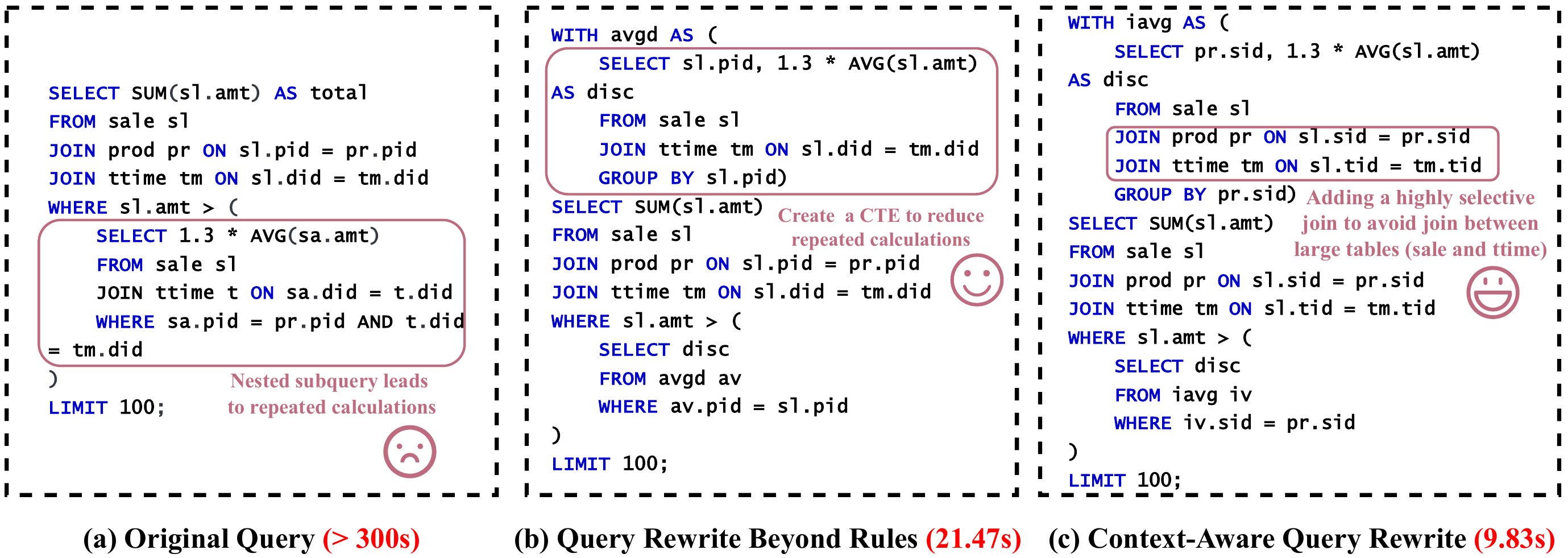}
    \vspace{-1em}
    \caption{An Example of Using Strategies Beyond Rewrite Rules (Example \ref{example:1}) and Using Context-Aware Analysis (Example \ref{example:2})}
    \label{fig:example}
    \vspace{-0.5em}
\end{figure*}
    
\begin{figure}[t]
    \centering
    \includegraphics[width=0.95\linewidth]{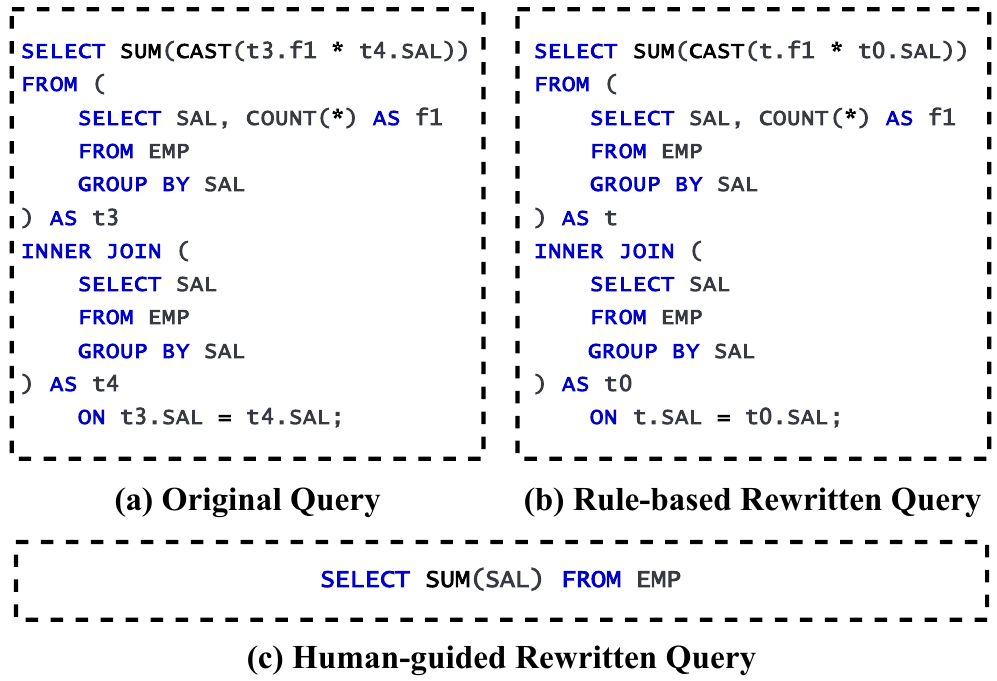}
    \vspace{-0.8em}
    \caption{An Example of Using Query Intent (Example \ref{example:3})}
    \label{fig:example1}
    \vspace{-1em}
\end{figure}

However, the above approaches 
can handle only a small subset of queries and may lead to performance regressions~\cite{lr,llm-r2,r-bot}. This limitation stems from the intrinsic challenges of rule-based query rewrite. First, the rich features of SQL and the nuances of semantics make it rather difficult to discover and verify new rewrite rules~\cite{extensible,hottsql,wetune}. Even state-of-the-art rule discovery system~\cite{wetune} can only support a narrow subset of operators. Second, fixed rewrite rules rely on pattern matching and therefore are fundamentally unable to optimize unseen or complex query patterns~\cite{Leis2015, querybooster, slabcity}. Third, many effective rewrite strategies cannot be expressed by rewrite rules~\cite{slabcity}. For example, the  Common Table Expressions (CTE) conversion in Example \ref{example:1} cannot be captured by rewrite rules. This is because such conversion requires explicit control over evaluation order and sub-plan sharing, which is beyond the scope of relational-algebra rewrite rules~\cite{graefe1993query,wetune,apache}.

Given these limitations, we revisit the query rewrite problem from the perspective of human experts and identify three key capabilities that humans exhibit, but rule-based systems lack. (1) Humans can perform rewrites that follow certain rules but exceed the expressiveness of standard pattern matching, and in some cases, go beyond what can be captured by formal rewrite rules (Example~\ref{example:1}). (2) Humans can conduct context-aware rewrites by considering the underlying data distributions and cost estimates provided by the query optimizer (Example~\ref{example:2}). (3) Experts can apply intent-oriented rewrites that capture the underlying goal of the query, which are not easily expressed as explicit rules (Example \ref{example:3}). 

\begin{example}
\label{example:1} \textit{Query Rewrite Beyond Rewrite Rules.} Figure \ref{fig:example} (a) is a simplified query from the DSB benchmark~\cite{ding2021dsb}. To avoid repeated calculation of nested subqueries,  Figure \ref{fig:example} (b) rewrites them using Common Table Expressions (CTEs). This enables the database engine to compute intermediate results once and reuse them, leading to a \textbf{37.5 $\times$} reduction in query latency in our experimental setup.
\end{example}

\begin{example}
\label{example:2}
    \textit{Context-Aware Query Rewrite.} In Figure \ref{fig:example} (c), instead of directly joining two large tables (the sale with 10 M rows and the ttime with 1.4 M rows), we first join the sale table with the prod table (100 K rows). This highly selective join would produce an intermediate result of only ~100 K rows. Next we join this with ttime table and yield a \textbf{2.79 $\times$} speedup over (b) in our experimental setup. Such rewrites, which reduce intermediate size, are rare in rule-based systems and require analyzing data distribution and selectivity.
\end{example}

\begin{example}
\label{example:3}
    \textit{Query Rewrite based on  Query Intent.} Figure \ref{fig:example1} (a) shows a real-world query from Apache Calcite~\cite{apache_calcite_test}. Rule-based rewrite cannot optimize this query (Figure~\ref{fig:example1} (b)). Identifying that the query intent is to sum salaries directly from the EMP table, humans can simplify the original query by eliminating unnecessary joins and sub-queries that do not contribute to the intent of the query, achieving a \textbf{2.17 $\times$} execution improvement in our experimental setup. 
\end{example}

Although human experts can produce significantly better SQL rewrites than rule-based methods, manual rewriting does not scale well, especially when there are millions of poorly written queries to be optimized in cloud environments. Recently, LLMs have demonstrated nearly human-level semantic and reasoning abilities, making them a natural choice for query rewrite. However, this task remains challenging due to the hallucination problem of LLMs: they may generate responses that are plausible but factually incorrect or semantically invalid. Specifically, there are two fundamental challenges. \textit {\textbf{C1. Ensuring Equivalent Rewrites.}} Producing equivalent rewrites requires a precise understanding of the rich features of SQL and the nuances of semantics. Even the most powerful LLMs produce rewrites with syntax or semantic errors in nearly 20\% of cases (Section \ref{sec:7}). \textit {\textbf{C2. Ensuring Optimized Rewrites.}} Determining whether a rewritten query improves performance depends on multiple factors, including query intent, data distribution, optimizer estimates, system states, and others. These contextual factors are typically unavailable to LLMs, and enabling LLMs to effectively utilize such information remains a non-trivial problem~\cite{gptuner,bai-etal-2024-longbench,llmqo}.

To address the limitations of rule-based query rewrite, the poor scalability of manual rewrite, and the unreliable nature of LLMs, we propose \system, a training-free and feedback-aware system that smartly leverages LLMs to rewrite SQL queries into semantically equivalent forms with significantly improved performance. \system\ supports a broader range of query patterns and rewrite strategies, while maintaining good scalability. 
It is primarily designed for long-running OLAP workloads, where execution time dominates and the cost of query rewriting is negligible. 
First, to enable LLMs to capture query data characteristics and leverage database feedback, we design a multi-agent framework controlled by a finite state machine (FSM) to iteratively refine the rewrite process. We decompose the complex rewrite process into a set of subtasks, each handled by a specialized LLM agent to reduce hallucinations and improve rewrite ability (addressing \textbf{C1, C2}). Second, to enhance the ability of LLMs to generate high-quality query rewrites, we provide a rewrite middleware that includes: (1) a structured knowledge base extracted from official documentation and query rewrite discussions on web forums,
(2) a hybrid SQL corrector that combines tool-based and LLM-based equivalence checking to ensure query equivalence,
Third, we employ a novel hint injection technique to generate better execution plans for rewritten queries. This ensures that query execution exactly follows the optimized plan recommended by LLM agents, avoiding unintended modifications by the query optimizer due to inaccurate cost estimations. Additionally, this technique enables further improvements by incorporating physical-level optimizations (addressing \textbf{C2}).

We extensively evaluate \system\ against state-of-the-art query rewrite systems on widely used benchmarks (e.g., TPC-H, DSB, and Calcite), which shows that our system achieves up to a \textbf{21.9\%} speedup in query execution performance over the \textbf{best-performing} alternative. Moreover, it delivers \textbf{24.1\%} additional rewrites compared to that alternative, covering query cases that previous systems could not handle.
In summary, we make the following contributions:

\begin{itemize}[leftmargin=*, noitemsep, topsep=0pt] 
\item  We propose \system, a training-free and feedback-aware query rewrite system that leverages LLM agents to support a broader range of query patterns and rewrite strategies than existing methods (Section \ref{sec:3}).
\item  We design a multi-agent framework controlled by a finite state machine to enable LLM agents to use external tools and enhance the rewrite process with database feedback (Section \ref{sec:5}).
\item  We develop a rewrite middleware including a structured knowledge base, a hybrid SQL corrector, and an agent memory buffer to enhance the rewrite capability of LLM agents (Section \ref{sec:4}).
\item  We propose a hint injection technique to produce better execution plans for rewritten queries. (Section \ref{sec:6})
\item  We conduct extensive experiments to demonstrate that our system significantly surpasses state-of-the-art methods in both query performance and query coverage (Section \ref{sec:7}).   
\end{itemize}

\section{Preliminaries and Related Work}
\label{sec:related work}
In this section, we first formally define the query rewriting problem and then discuss related work on query hints and LLMs.

\begin{figure*}[ht]
    \centering
    \includegraphics[width=\linewidth]{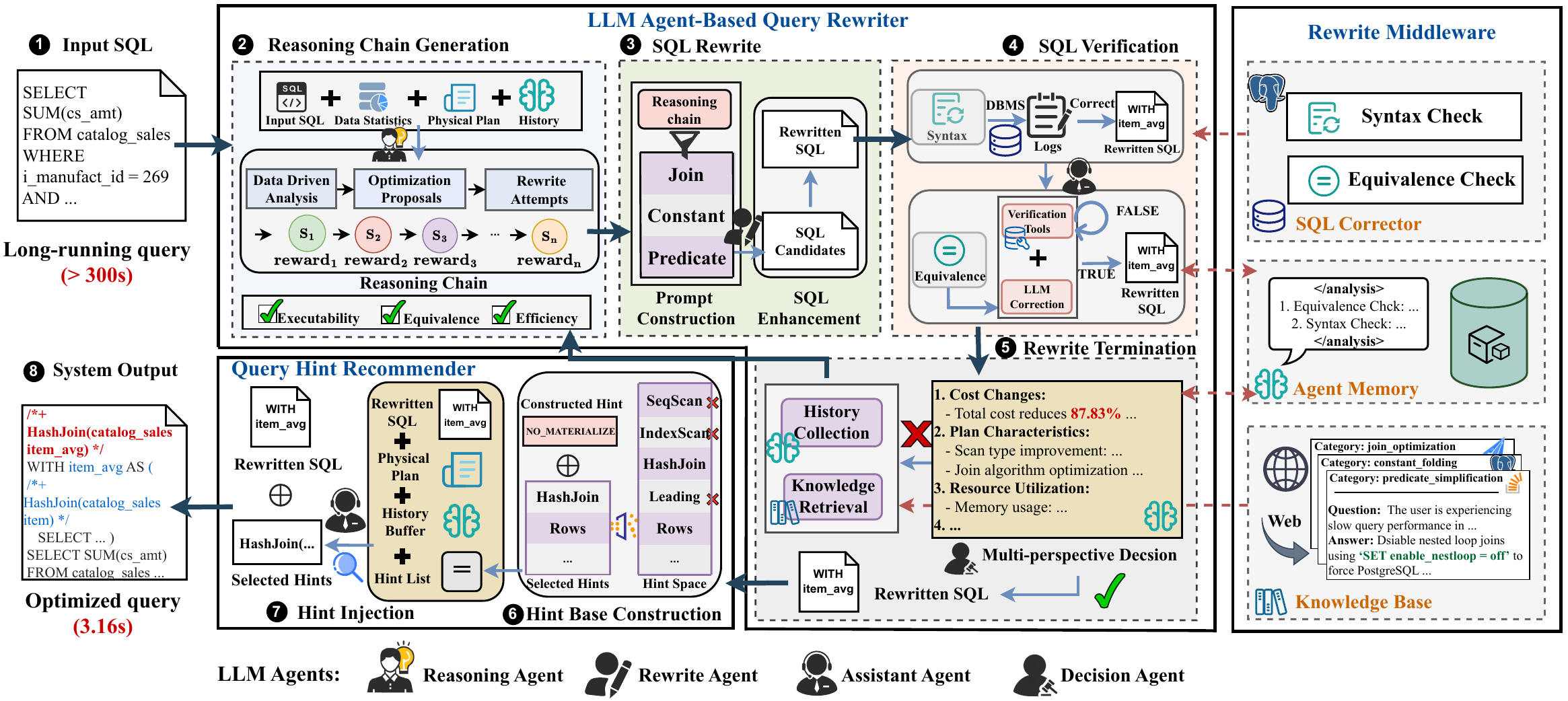} 
    \vspace{-2em} 
    \caption{System Overview of \system}
    \label{fig:overview}
    \vspace{-1em}
\end{figure*}

\subsection{Query Rewrite}
\label{sec:2.1}
We first define query rewrite in the realm of database systems:

\begin{definition}[Query Rewrite]
Query rewrite is a process of transforming a query into another query that is \textit{semantically equivalent} but with improved performance. It is a preliminary step prior to query optimization, operating at the level of the \textit{user-accessible application programming interface (API)} (e.g., SQL).
\label{def: qr}
\end{definition}
\vspace{-0.5em}
\begin{definition}[Query Equivalence]
Two queries are considered \textit{equivalent} if they produce the same result for any valid instance of the database schema.
\end{definition}

Conventional studies treat query rewrite as a transformation between plain SQL statements, typically using rule-based methods. There are two lines of work:

\noindent (1) \textit{Discovery of new rules}. WeTune~\cite{wetune} automatically generates and verifies logical plan transformations, but it handles only a limited set of operator types and algebraic rules. Equality saturation techniques~\cite{10.1145/3485496} and user-guided systems like QueryBooster~\cite{querybooster} still require manual intervention and cover only narrow classes of queries. More recently, GenRewrite~\cite{genrewrite} leverages LLMs to describe rewrite rules in natural language (NLR2s), enabling more flexible rule specification. In practice, its effectiveness depends on the coverage of available equivalence checks, and validating a large number of candidate rewrites through execution can introduce substantial overhead~\cite{genrewrite,slabcity}.


\noindent (2) \textit{Effective use of existing rules}.
LearnedRewrite~\cite{lr,lr_demo} uses Monte Carlo Tree Search (MCTS) with learned cost models to explore the space of rule applications, but its performance depends on the accuracy of cost model estimation. 
LLM-R$^2$~\cite{llm-r2} retrieves similar past rewrites using a pre-trained selector and employs them as in-context examples for rule selection. 
R-Bot~\cite{r-bot} adopts retrieval-augmented generation to recommend rewrite rules through evidence retrieval and step-by-step reasoning.
These approaches largely retrieve and apply rewrite knowledge based on query structure and then select from predefined rule sets. Consequently, they may not fully capture dataset-dependent factors such as database scale or column statistics, and their rule-centric formulation can miss effective expert strategies that are hard to express as explicit rewrite rules.


\subsection{Query Hints}
\label{sec:2.2}
Query hints are SQL extensions that provide instructions to the database's query engine to influence the selection of execution plans~\cite{bao,autosteer,fastgres,steeringqo}. These hints allow users to exert precise control over how the database optimizer chooses physical operators and access methods.
As shown below for \texttt{pg\_hint\_plan} extension~\cite{pg_hint_plan}, the hint \textit{/*+ HashJoin(employees departments) */}  forces the query engine to use a hash join for the $employees \bowtie departments$ operation. This hint can override the optimizer’s default cost-based plan selection.

\begin{lstlisting}[language=SQL, 
    xleftmargin=15pt, xrightmargin=10pt, 
    framexleftmargin=5pt, framexrightmargin=5pt, 
    rulesep=5pt, numbers=none]
/*+ HashJoin(employees departments) */
SELECT *
FROM employees e
JOIN departments d ON e.dept_id = d.id
WHERE e.salary > 50000;
\end{lstlisting}


Prior works on query hint selection~\cite{bao,autosteer,fastgres} apply boolean hints to entire queries, which often leads to suboptimal performance because different operators within a query may require different hints. Proto-X~\cite{zhang2024holon} improves upon this by assigning hints at the operator level. However, it operates in an offline setting that requires exploring multiple configurations and repeatedly executing the complete workload, whereas query rewriting does not require repeated workload execution. UniTune~\cite{zhang2023unified} supports query rewriting through LearnedRewrite~\cite{lr} but does not incorporate hints. In contrast, \system\ employs \texttt{pg\_hint\_plan} to enable fine-grained, operator-level control (e.g., selection of join algorithms and adjustment of cardinalities) and jointly optimizes both query rewriting and hint selection.

\subsection{LLM and LLM Agents}
\label{sec:2.3} 
\noindent \textbf{LLM for Database Tasks.} LLMs are increasingly applied to database tasks text-to-SQL~\cite{text2sql2, codes, omnisql, xiyansql}, SQL workload generation~\cite{SQLBarber, lao2025sqlbarberleveraginglargelanguage}, and database optimization~\cite{gptuner,llmidx,gptuner-demo,llmqo,gptuner-record}, with recent work focusing on query optimization~\cite{llm-r2,r-bot,genrewrite}. Although these methods show that LLMs can understand and modify SQL, the inherent complexity of query rewrite presents substantial challenges.

\noindent \textbf{LLM Agents.} The primary advantage of LLM agents over LLMs is their ability to collect feedback from the execution environment (e.g., database execution metrics) and adapt their subsequent reasoning and outputs accordingly~\cite{wu2025agenticreasoningreasoningllms}.
They have proven particularly effective in code generation~\cite{liu2023your,yang2024sweagentagentcomputerinterfacesenable},  automated diagnostics~\cite{d-bot,mdagent,thompson2023large}, and knowledge management~\cite{d-bot,asai2024self}.
Moreover, collaborative deployments of multiple LLM agents often outperform single‑agent systems on complex reasoning and planning tasks~\cite{multi1,multi2,multi3,multi4}.
\system\ applies this multi-agent methodology to query rewrite by gathering DBMS feedback and adjusting each query candidate accordingly.

\section{System Overview}
\label{sec:3}
We present the architecture of \system\, as shown in Figure~\ref{fig:overview}. It comprises three components: LLM Agent-Based Query Rewriter, Rewrite Middleware, and Query Hint Recommender.

\label{sec:3.3}



\noindent \textbf{LLM Agent-Based Query Rewriter.} 
The rewriter uses a training-free and feedback-aware LLM agent-based FSM to rewrite queries in multiple stages.
First, \ding{182} the user provides the input SQL along with the corresponding database configuration (e.g., DBMS type, host address). Second, \ding{183} the rewriter obtains metadata (e.g., database statistics, query information) to generate a reasoning chain, which includes rewrite proposals and rewritten SQL candidates. This chain is then used to generate the rewritten SQL for further enhancement and validation. Next, \ding{184} we consolidate the SQL candidates to construct the LLM prompt, and apply global SQL enhancements to find the optimal rewritten SQL. Finally, \ding{185} we validate the syntax correctness and semantic equivalence of the rewritten SQL, and \ding{186} verify the correctness and effectiveness of the rewritten query to decide whether to proceed to Step \ding{187} or further rewrite this query based on the rewrite history.




\noindent \textbf{Rewrite Middleware.}
The toolkit provides three specialized tools to boost LLM rewrite ability. First, we construct a structured knowledge base using a multi-source integration to guide high-quality rewrite proposals (Section~\ref{sec:5.1}). 
Second, a hybrid SQL corrector then validates the syntactical correctness and semantic equivalence of each candidate rewrite, using rewrite history to trigger further refinement when necessary. (Section~\ref{sec:5.2}). Third, we design an agent memory buffer that captures key context from previous interactions to reduce communication overhead and LLM hallucination (Section \ref{sec: 5.3}).

\noindent \textbf{Query Hint Recommender.} The recommender explores the query hints in two steps. \ding{187} First, we construct a promising query hint base, thereby reducing the hint search space and enabling faster, more accurate hint selection (Section \ref{sec:6.1}). \ding{188} Second, we leverage an LLM agent with data statistics to select appropriate fine-grained hints for the rewritten query to improve its execution performance (Section \ref{sec:6.2}). 
\ding{189} Finally, the selected hints are injected into the rewritten SQL to produce the final output.

\begin{figure}[t]
    \centering
    \includegraphics[width=1.0\linewidth]{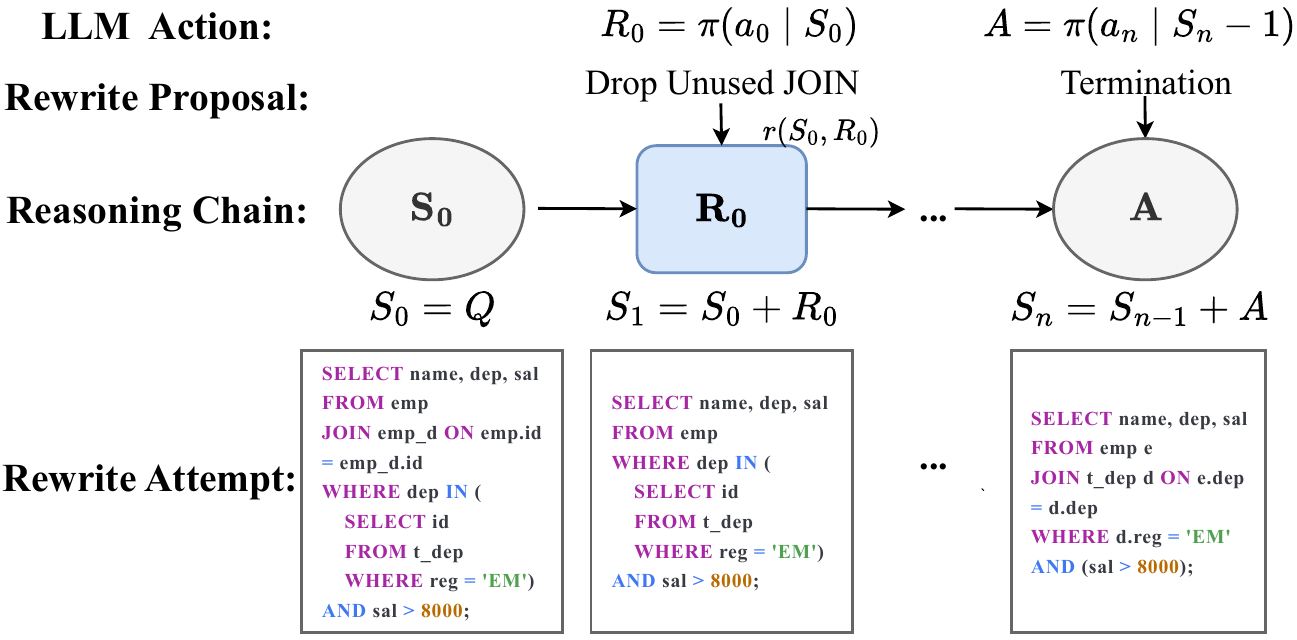}
    \caption{Reasoning Chain of MDP-based Reasoning Agent}
    \label{fig:MDP}
    \vspace{-1em}
\end{figure}

\section{LLM Agent-Based Query Rewriter}
\label{sec:4}

In this section, we present our LLM agent-based query rewriter and describe how to construct an efficient, robust query‐rewriting workflow using LLM agents. 

\subsection{MDP-based Reasoning Agent}
\label{sec:4.1}
Query rewrite is fundamentally a sequential decision problem~\cite{131472, lr}, each transformation step changes the query structure and limits future optimization options. Reasoning LLM agents (e.g., DeepSeek-R1~\cite{DeepSeek-r1}), capable of chain-of-thought planning and multi-step logical inference, appear well-suited to this task. However, our experiments show that they still produce 13.0\% nonequivalent outputs and yield minimal execution improvements (Section \ref{sec:7.2}). We observe that query rewrite mirrors a Markov Decision Process (MDP), where states are query forms, actions are rewrite steps, and rewards are cost reductions~\cite{mdp,liu2023reason,tutorial}. By framing rewrite in this way, we provide the reasoning LLM agent with a clear, stepwise optimization goal that guides each inference step. Therefore, we design an MDP-based reasoning agent to produce high-performing rewrites.

We model the query rewrite process as a Markov Decision Process $(\mathcal{S},\mathcal{A},\mathcal{T},r,\gamma)$. Starting from the original query \(Q\), each \emph{state} $S_t\in\mathcal{S}$ represents the current form of the query with any semantic-level rewrite attempts.
At each timestep $t$, the agent chooses either a \emph{refinement action} $R_t$ (e.g., join reordering, predicate pushdown) or a \emph{terminal action} $\mathcal{A}$ that emits the final rewritten query.  This yields the deterministic transition:
\[
S_{t+1} = 
\begin{cases}
\mathcal{T}(S_t, R_t) = S_t + R_t, & \text{if a refinement is applied},\\
\mathcal{T}(S_t, a)   = S_t + \mathcal{A},   & \text{if rewriting terminates}.
\end{cases}
\]

To guide the search toward efficient rewritten queries, we define a domain‐specific \emph{reward function} once $S_t$ transitions to $S_{t+1}$ under a refinement action $R_t$:
\[
r(S_t, R_t) \;=\; \mathrm{Cost}(S_t)\;-\;\mathrm{Cost}(S_{t+1}),
\]
where $\mathrm{Cost}(\cdot)$ denotes the DBMS optimizer’s estimated execution cost (via \texttt{EXPLAIN}) and LLM's evaluation. Intuitively, each positive reward corresponds to a drop in estimated cost. The agent’s objective is captured by the \emph{action‐value function} \(Q^\pi\) under policy \(\pi_\theta(a \mid S_t)\), where action $a \in R_t$ :
\[
Q^\pi(S_t, a_t)
\;=\;
\mathbb{E}_{\pi}\Bigl[\sum_{k=0}^{T-t}\gamma^k\,r(S_{t+k},a_{t+k})
\;\Bigm|\;S_t,a_t\Bigr],
\]
which represents the expected total discounted reward starting from state \(S_t\) after taking action \(a_t\).  The \emph{optimal} action‐value function \(Q^*\) then satisfies the Bellman equation:
\[
Q^*(S_t,a_t)
\;=\;
r(S_t,a_t)
\;+\;
\gamma\,\max_{a'\in\mathcal{A}}Q^*(S_{t+1},a'),
\]

In practice, we realize the policy $\pi_\theta(a\mid S_t)$ as a reasoning‐specialized LLM (e.g., DeepSeek-R1) that generates each action conditioned on the current query state.

As shown in Figure \ref{fig:MDP}, the agent transitions from the initial state \(S_0\) through a sequence of refinement actions \(\{R_0, R_1, \dots\}\), and finally applies the terminal action \(A\) to emit the final rewritten query.
Concretely, at each timestep \(t\), the agent is prompted with the current SQL form \(S_t\) and analyzes the query structure via its internal chain‐of‐thought, scores candidate refinements based on the expected cost reduction, and samples the highest‐scoring \(R_t\).

\subsection{FSM-based Query Rewrite}
\label{sec:4.2}
Single-agent approaches often produce suboptimal or invalid rewrites because a single agent is responsible for managing the entire rewrite pipeline (e.g., propose rewrites, verify validity, and assess efficiency), which overloads the model’s context capacity and ultimately increases the risk of hallucination.
Inspired by multi-agent frameworks that assign specialized agents to distinct tasks~\cite{metagpt}, we break down query rewrite into discrete stages, each of which is handled by a dedicated LLM agent. To ensure the multi-agent system's stability and correctness~\cite{multi-agent-chanllenges}, we implement a Finite State Machine (FSM) to orchestrate stable stage transitions.



\subsubsection{Query  Rewrite Process Decomposition}  
Given an environment $\mathcal{E}$ representing a database system, an \emph{LLM Agent Group for Query Rewrite} is a tuple $\mathcal{A} = (\mathcal{G}, \mathcal{M}, \mathcal{F}, \mathcal{R})$ where:
\begin{itemize}[leftmargin=*, noitemsep, topsep=0pt]
    \item $\mathcal{G} = \{A_1,...,A_n\}$ is a set of cooperative LLM-based agents, each with specialized capabilities $C_i$ for tasks.
    
    \item $\mathcal{M}: \mathcal{Q} \times \mathcal{E} \rightarrow \mathcal{S}$ is a transition that breaks down the input query $Q \in \mathcal{Q}$ into intermediate states $s \in \mathcal{S}$.
    
    \item $\mathcal{F}: \mathcal{S} \times \mathcal{H} \rightarrow \mathcal{Q}'$ is a transformation function that produces rewritten queries $Q' \in \mathcal{Q}'$ based on the current state and historical feedback $h \in \mathcal{H}$ from $\mathcal{E}$.
    
    \item $\mathcal{R}: \mathcal{Q}' \times \mathcal{E} \rightarrow \mathbb{R}^+$ is a reward function that evaluates rewrite quality through execution metrics and guides iterative refinement.
\end{itemize}

Specifically, the Agent takes action $O_i$ in state $s_i$, the environment provides feedback $O_i'$, and the Agent updates its state and decides the next action based on the feedback as:
\[
\mathcal{A}_{s_i} \xrightarrow{O_i} \mathcal{E} \xrightarrow{O_i'} \mathcal{A}_{s_{i+1}}
\]
where $\mathcal{A}_{s_i}$ denotes the Agent in state $s_i$, and $\mathcal{A}_{s_{i+1}}$ denotes the Agent in state $s_{i+1}$. 

\begin{figure}[t]
    \centering
    \includegraphics[width=1.0\linewidth]{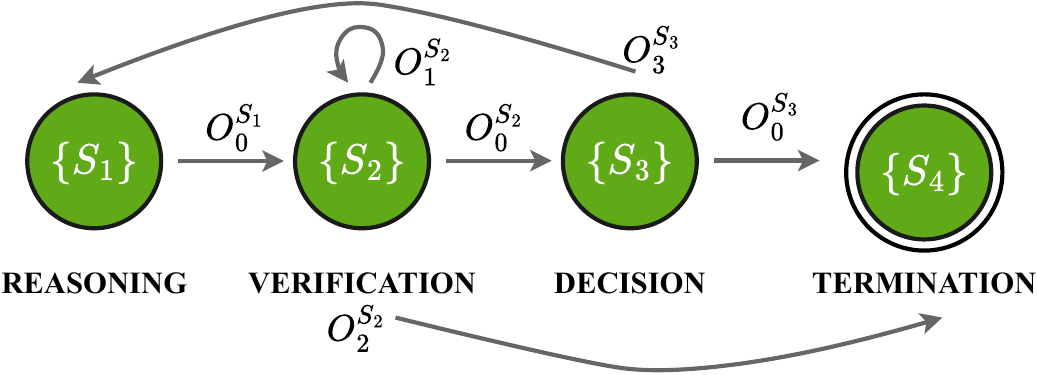}
    \vspace{-2em}
    \caption{The Workflow of FSM-based Query Rewrite}
    \label{fig:fsm}
\end{figure}

In practice, we decompose the rewrite process into four stages: \textit{Reasoning}, \textit{Verification}, \textit{Decision} and \textit{Termination}, as illustrated in Figure~\ref{fig:fsm}. The \textit{Reasoning} stage generates candidate SQL rewrites. The \textit{Verification} stage validates syntax and semantic equivalence, correcting any errors identified. The \textit{Decision} stage determines whether to continue or terminate the rewrite process. The \textit{Termination} stage ends the rewrite process and outputs the rewritten SQL with rewrite proposals.


\subsubsection{Specialized LLM Agents} As shown in Table~\ref{tab:agent-roles}, we deploy three specialized LLM agents alongside the core \textit{MDP-based Reasoning Agent} in Section \ref{sec:4.1} to drive the rewrite pipeline for the defined FSM stages. Then we propose an LLM agent-based query rewrite methodology detailed in Algorithm \ref{algo:_fsm_llm}:
\begin{itemize}[leftmargin=*,noitemsep]
  \item \textbf{Reasoning stage} (\(S_1\)): In this stage, the \textit{MDP-based Reasoning Agent} generates a reasoning chain, containing detailed rewrite proposals and SQL candidates. The \textit{Rewrite Agent} then serves as a reward model, extracting and refining SQL candidates from the reasoning chain (Lines 3-7).
  \item \textbf{Verification stage} (\(S_2\)): In this stage, the \textit{Assistant Agent} validates syntactical correctness and semantic equivalence by using the hybrid SQL corrector (Lines 8-12).
  \item \textbf{Decision stage} (\(S_3\)): In this stage, the \textit{Decision Agent} reassesses the rewritten SQL's efficiency with a comprehensive report and determines whether the rewrite process should continue or terminate (Lines 13-19).
\end{itemize}

\begin{table}[t]
  \centering
  \caption{Specialized Agents in the Finite State Machine} 
  \label{tab:agent-roles}
  \begin{tabular}{|c|c|c|}
    \hline
    State & Agent & Agent Role \\ 
    \hline
    $S_1$    &  \makecell{MDP-based\\Reasoning Agent}  & Generate Reasoning Chain          \\ 
    \hline
    $S_1$    & Rewrite Agent     & Extract \& Refine SQL candidates           \\ 
    \hline
    $S_2$ & Assistant Agent   & Validate Syntax \& Equivalence              \\ 
    \hline
    $S_3$     & Decision Agent    & Score \& Control Termination\\ 
    \hline
  \end{tabular}
\end{table}

\begin{algorithm}[t]
\SetAlgoLined
\caption{LLM-based Query Rewrite Algorithm}
\label{algo:_fsm_llm}
\SetKwInput{Input}{Input}
\SetKwInput{Output}{Output}

\Input{Original SQL Query $Q_0$; LLM Agent $\mathcal{M}$; Tool Repository $T$; Knowledge Base $\mathcal{K}$}
\Output{Optimized SQL Query $Q^*$ with reduced cost}

Initialize State $S\gets\textsc{Reasoning}$; Iteration $t\gets 0$; Advanced Knowledge $\advice \gets\emptyset$\;

\While{$S\neq\textsc{Termination} \land t\leq T_{\max}$}{
    \uIf{$S=\textsc{Reasoning}$}{
        $\mathcal{C}\gets\text{Generate-Reasoning-Chain}(\mathcal{M}, Q_0, T)$\;
        $\langle Q^P, \advice\rangle\gets\text{Extract-Rewrite-Proposal}(\mathcal{C})$\;
        $\langle Q^E,\text{EarlyStop}\rangle\gets\text{Enhance-Query}(\mathcal{M}, Q^P, \advice)$\;
        $S\gets \text{EarlyStop} \vspace{0.25pt}? \vspace{0.25pt} \textsc{Termination} : \textsc{Verification}$\;
    }
    \uElseIf{$S=\textsc{Verification}$}{
        \While{$Q^E$ not pass verification}{
            Correct syntax and equivalence of $Q^E$ via $T$\;
        }
        $S\gets\textsc{Decision}$\;
    }
    \uElseIf{$S=\textsc{Decision}$}{
        $\mathcal{R}\gets\text{Generate-Report}(Q^E,Q_0,\mathcal{M},T)$\;
        \uIf{$\text{Cost-Check}(Q^E,\mathcal{R},\mathcal{M})$}{
            $Q^*\gets Q^E$; $S\gets\textsc{Termination}$\;
        }
        \uElse{
            $\advice\gets\advice\cup\text{Retrieve-Knowledge}(\mathcal{K},Q^E)$\;
            $S\gets\textsc{Reasoning}$\;
        }
    }
    $t\gets t+1$\;
}
\Return $\langle Q^*, \text{cost}(Q^*), \advice\rangle$\;

\end{algorithm}
 \vspace{-1em}
 
\noindent \textbf{Rewrite Agent.} After the reasoning agent generates a chain, the rewrite agent extracts proposals from each node, groups them into four categories matching our structured knowledge base, evaluates each proposal’s expected reward, and selects the top SQL candidate. Since this candidate may introduce new structural complexities, it cannot be used directly for verification. Therefore, the rewrite agent undertakes an SQL enhancement process to uncover additional rewrite opportunities. 



This template provides a concise, user-curated checklist of the effective SQL enhancement transformations. Guided by this checklist, the rewrite agent revisits the selected SQL candidate to identify any remaining optimization opportunities. If none are found, it returns the original query unchanged.

\noindent \textbf{Assistant Agent.}  
The assistant agent uses the hybrid SQL corrector (Section~\ref{sec:5.2}) to ensure the syntactic correctness and semantically equivalence of queries. This is performed in two steps:
\begin{itemize}[leftmargin=*, noitemsep, topsep=0pt]
    \item \textit{Syntax correction.} If syntax errors occur, it attempts up to \(K_{\max}\) repairs.  If correction still fails, the current FSM iteration is aborted and control returns to the reasoning stage.
    \item \textit{Equivalence correction.} Once the syntax error is corrected, the agent submits the query pair \(\langle Q_{\text{orig}},Q_{\text{new}}\rangle\) to verify the equivalence and correct nonequivalent queries.
\end{itemize}

This two-phase process ensures that only valid, semantically equivalent rewrites advance to the Decision stage. Although equivalence verification may return the original input SQL, it avoids the far greater cost of returning nonequivalent SQL while preventing performance degradation from excessive loop iterations.

\noindent \textbf{Decision Agent.} The decision agent implements a cost-aware decision mechanism to determine whether the rewrite process should terminate within budget constraints. It contains two steps: 

\textit{(i) Generate Comprehensive Report.} The decision agent analyzes the rewritten query across four dimensions, including cost changes, plan characteristics, resource utilization, and other improvements. This multi-dimensional approach overcomes the limitations of relying solely on EXPLAIN cost estimates, preventing from accepting misleading or suboptimal rewrites.

\textit{(2) Evaluation and Feedback Loop.} Based on the generated report, the decision agent makes a binary decision to decide whether the rewrite process should terminate or not. If \textit{True}, the Decision agent terminates the FSM process; if \textit{False}, it retrieves higher-confidence rewrite proposals from the knowledge base, then stores the current query, report, and retrieved knowledge in an agent shared buffer for the next FSM iteration. We deliberately avoid introducing knowledge base retrieval in the initial state, because reasoning models have input content limitations—excessive guidance information can reduce the generation freedom, potentially compromising the quality of reasoning proposal chain~\cite{DeepSeek-r1}. Finally, the FSM enters the Termination stage and outputs the final rewritten query along with its comprehensive optimization report.

\section{Rewrite Middleware}
\label{sec:5}
In this section, we introduce the Rewrite Middleware. It comprises three key components: a structured knowledge base of query rewrite strategies (Section \ref{sec:5.1}), a hybrid SQL corrector ensuring syntactic validity and semantic equivalence (Section \ref{sec:5.2}) and an agent memory buffer for agent context management (Section \ref{sec: 5.3}).

\subsection{Structured Knowledge Base}
\label{sec:5.1}

Many query rewrite strategies developed by DBAs cannot be effectively utilized by LLMs. This limitation arises because such knowledge is sparsely represented in the LLMs’ training data, making it difficult for them to extract and apply these strategies without explicit guidance provided in the prompt. To address this, we construct a structured knowledge base to systematically manage various rewrite strategies. The construction process consists of four steps:
(i) collecting SQL rewrite knowledge from diverse sources, (ii) filtering unreliable content, (iii) enriching the collected entries, and (iv) retrieving the most relevant rules for each rewrite. The Knowledge Base $\mathcal{K}$ is defined as follows:



\textit{Definition 1 (Knowledge Base).} A \textit{Knowledge Base} $\mathcal{K}$ is a set of question and answer (Q\&A) pairs, denoted as $\mathcal{K}=\{\langle q_1,a_1 \rangle, \langle q_2,a_2 \rangle, \allowbreak ..., \langle q_n,a_n \rangle \}$. Each Q\&A pair $\langle q_i,a_i \rangle$ includes: 
\begin{itemize}[leftmargin=*, noitemsep, topsep=0pt]
    \item A question $q_i= \langle Text_{que_i},SQL_{que_i} \rangle$, where $Text_{que_i}$ describes a specific SQL rewrite issue in natural language, and $SQL_{que_i}$ is the original SQL query that the system needs to interpret or optimize.
    \item An answer $a_i=\langle Text_{ans_i},SQL_{ans_i}\rangle$, where $Text_{ans_i}$ explains the strategies used to rewrite $SQL_{que_i}$, and $SQL_{ans_i}$ is the rewritten SQL query.
\end{itemize}

\noindent \textit{Step 1: Collecting rewrite knowledge.}
Our knowledge is primarily sourced from official database documentation and the database community. Since rewrite strategies are often indirectly presented and scattered throughout official documents~\cite{postgresqldocs,mysql_docs,sqlserver_docs,oracle_db_docs}, we employ LLMs to identify and summarize them into key points. We also collect knowledge from Stack Overflow~\cite{stackoverflow}, as it has abundant real-world query rewrite examples and insights accumulated over decades. We retrieve all Q \& A units tagged with ``query rewrite'' or ``query optimization'' that include at least one answer. In total, we collected 3,432 valid Q \& A units.

\noindent \textit{Step 2: Filtering untrustworthy content.}
We begin by filtering out answers that receive more dislikes than likes. Next, we prioritize solutions achieving consensus within the discussion, using an LLM to determine if consensus has been reached. Following this, we employ a majority vote by LLMs to identify the most effective solution from these consensus-based responses. This process yields a total of 241 high-quality Q\&A units.

\noindent \textit{Step 3: Enhancing the knowledge.}
An LLM first summarizes questions and user's responses from website Q\&A units. These summaries are then augmented with corroborating information from official documents. To do this, we embed user response summaries using Sentence Transformers~\cite{sentence-transformer}. These embeddings are then used to query a pre-indexed knowledge base of extracted points, retrieving the top three matches via cosine similarity. Finally, an LLM evaluates these matches to confirm that they effectively explain the rewrite strategy underlying the Q\&A unit. Suitable matches are incorporated, enriching Q\&A units with concise explanations that aid future retrieval.

\begin{table}[!t]
\centering
\caption{Distribution of Knowledge UNits Across Categories}
\vspace{-0.5em}
\begin{tabular}{|m{1.22cm}|m{1.42cm}|m{1.42cm}|m{1.42cm}|m{1.2cm}|}
\hline
\centering\textbf{Join} &
\centering\textbf{Constant} &
\centering\textbf{Predicate} &
\centering\textbf{CTE} &
\centering\textbf{Others} \tabularnewline
\hline
\centering 122 & \centering 2 & \centering 21 & \centering 20 & \centering 76 \tabularnewline
\hline
\end{tabular}
\label{tab:kb-stats}
\vspace{-4em}
\end{table}

\noindent \textit{Step 4: Retrieving relevant knowledge.}
We classify these 241 high-quality units into five
categories: predicate,  constant, join, CTE and others, which capture the essential aspects of query rewrite and provide a foundation for precise retrieval (Table~\ref{tab:kb-stats}).s
Since knowledge is often context-specific (i.e., most rewrite strategies depend heavily on factors such as
schema definitions, data distribution and available 
indexes), its effectiveness may diminish under changing conditions. 
In contrast, context-independent knowledge is relatively rare and involves general-purpose optimizations such as predicate simplification. To identify the most relevant 
knowledge, we use the BM25 algorithm~\cite{bm25} to retrieve the most relevant documents by computing the document relevance.

\begin{table*}[h]
\centering
\caption{Constructed Hint Base} 
\vspace{-1em}
\begin{tabular}{|l|l|l|}
\hline
\textbf{Query Hint  }          &\textbf{Grammar}                                      & \textbf{Description }                               \\ \hline
Hash Join                & \texttt{/*+ HashJoin(table table[ table...]) */}    & Forces the use of hash joins for specified tables. \\ \hline
No Nested Loop Join           & \texttt{/*+ NoNestLoop(table table[ table...]) */}    & Prevents the use of nested loop joins for specified tables. \\ \hline
No Merge Join                  & \texttt{/*+ NoMergeJoin(table table[ table...]) */}   & Prevents the use of merge joins for specified tables. \\ \hline
Row Correction       & \texttt{/*+ Rows(a b \#10) */}                        & Sets the number of rows for the join result.         \\ \hline
NOT\_MATERIALIZE            & \texttt{/*+ NOT\_MATERIALIZE(table) */}                & Inlines the CTE to avoid materialization overhead.   \\ \hline
\end{tabular}
\label{tab:hintbase}
\vspace{-1em}
\end{table*}

\subsection{Hybrid SQL Corrector}
\label{sec:5.2}
Hallucinations in LLMs can cause them to generate faulty queries, with two primary types of errors:
(1) \textit{Syntax Error}, where a generated SQL violates language grammar and cannot execute; 
(2) \textit{Equivalence Error}, where a generated SQL executes but yields results or semantics different from the original SQL.
Our proposed hybrid correction method targets both issues.

\noindent \textbf{Syntax Error Correction.}
Syntax errors, typically minor issues like misspelled keywords or incorrect column/table names, prevent the database from parsing the SQL query. An LLM, leveraging its understanding of SQL syntax and guided by error messages from the DBMS, can efficiently identify and fix these structural flaws. Because such corrections usually involve minimal changes targeting obvious mistakes, the query's original intended semantics are generally preserved with low overhead during this process.

\noindent \textbf{Equivalence Error Correction.}
Verifying the equivalence between any pair of SQL queries is known to be an NP-hard challenge~\cite{eq_np}, and whether this problem admits a polynomial-time solution remains an open question.
While numerous verification methods~\cite{cosette, wetune,qed, sqlsolver} rooted in algebraic or symbolic reasoning have been proposed to prove equivalence, they typically operate on constrained query sets (e.g., specific operators, advanced SQL constructs). 
Recent studies have highlighted the potential of LLMs in deducing query equivalence with high confidence~\cite{singh2024exploring,LLMSQLSolverCL}. However, relying solely on LLMs carries inherent risks. 
Therefore, we integrate traditional verification tools and LLM capabilities in a two-stage process, forming a hybrid approach that achieves broader and more reliable equivalence judgments.

\begin{itemize}[leftmargin=*, noitemsep, topsep=0pt]
    \item \textbf{Tool-based Verification.} We first use SQLSolver~\cite{sqlsolver}, an advanced verification tool, for an initial check. Given a pair of queries, $\langle Q_{orig}, \allowbreak Q_{rewr} \rangle$, SQLSolver determines their equivalence as either equivalent, nonequivalent, or unknown (due to timeout or other limitations). For query pairs where SQLSolver's outcome is unknown, verification proceeds to a subsequent stage, utilizing the proposed LLM-based approach.
    \item \textbf{LLM-based Verification and Correction.} Leveraging their ability to understand semantic relationships, LLMs can effectively check SQL query equivalence and correct nonequivalent rewrites. However, directly applying LLMs faces challenges, as they may be unsure for some complex queries or incorrectly identify a pair of nonequivalent queries as equivalent. To mitigate these issues, we employ an iterative verification and correction process. In this process, the LLM repeatedly compares the original query $Q_{orig}$ with the rewrite candidate, systematically identifying discrepancies and applying refinements. This iterative loop continues until the LLM confidently establishes query equivalence or a predefined time budget is exhausted. If the time budget is reached without the LLM confidently establishing equivalence, the system defaults to returning the original query ($Q_{rewr} = Q_{orig}$), ensuring that system performance is not compromised by unresolved complex rewrites. Our experiments on diverse workloads show that our method achieves higher rewrite equivalence rates than baselines (Figure \ref{fig:benchmarks}).
\end{itemize}

\subsection{Agent Memory Buffer}
\label{sec: 5.3}
Effective context sharing is crucial in LLM-based workflows. Simply logging the entire prompt history is problematic: 
\textit{(1) Hallucinations due to redundant context.} Lengthy SQL statements and redundant reasoning tokens can distract LLM, causing it to generate incorrect content~\cite{shi2023large}.
\textit{(2) Performance degradation under long-context overhead.} Unbounded history growth incurs quadratic attention costs and increases inference latency~\cite{jiang2023longllmlingua,kuratov2024babilong}.

To address these, we propose an agent memory buffer $\mathcal{B}$, containing memory slices $\mathcal{B} = \{m_1, m_2,..., m_n\}$, where each slice $m_i$ is dedicated to a specific category of critical information (e.g., query information, rewrite proposals). 
By extracting only essential messages across the rewrite process, our buffer curbs hallucinations and bounds context growth, thereby improving both correctness and efficiency in single-agent SQL rewrite workflows.

\section{Query Hint Recommender}
\label{sec:6}
Query rewrite applies high-level transformations to give the DBMS optimizer a better starting point for plan generation. While this improved starting point is beneficial, the optimizer, which dictates the finer-grained execution operations, can still generate suboptimal plans. This sub-optimality arises because the optimizer's decisions rely on cost models that are often inaccurate due to outdated statistics and simplifying assumptions (e.g., statistical independence of columns)~\cite{leo,neo,base}.
To exert more fine-grained control over plan selection, we propose embedding query hints~\cite{bao} within rewritten queries. In this work, we introduce a dedicated hint base and use LLMs to adaptively select and incorporate effective hints into rewritten queries.

\subsection{Hint Base Construction} 
\label{sec:6.1}
Most modern database systems (e.g., PostgreSQL, MySQL, Oracle) support query hints, either inherently or by extensions~\cite{pg_hint_plan, mysql_hints, oracle_hints}.  In this work, we use PostgreSQL’s \texttt{pg\_hint\_plan} extension~\cite{pg_hint_plan} to implement our query hint system. Our system supports a broader range of hints (e.g., disabling materialization and setting specific cardinality values),  whereas existing works~\cite{bao,autosteer,fastgres,steeringqo} are limited to Boolean hints.


\noindent \textbf{Selecting Existing Hints.} First, we exclude index-related and hardware acceleration hints as they are orthogonal to query rewrite optimization and can introduce hardware-specific dependencies. From the remaining hints, we employ an LLM with the prompt $p_{sel}$ = \textit{"Given a query hint set, you should select the most effective and practical hints for query rewrite and give the reasons."} The LLM's selections are then further refined based on expert review to ensure an optimal and practical set of hints.


\noindent \textbf{Introducing a Novel CTE Hint.} 
To provide more precise control over Common Table Expression (CTE) execution and associated performance issues, we introduce a new \texttt{NOT\_MATERIALIZE} hint. 
By default, PostgreSQL materializes CTEs~\cite{postgresqlcte}, which can lead to significant performance degradation, particularly when CTEs operate on small datasets or are executed infrequently~\cite{cte1,cte2}, due to the overhead of writing and reading intermediate results. 
Furthermore, the decision to materialize a CTE, as opposed to inlining it, directly impacts the optimizer's ability to perform effective predicate pushdown into subqueries, especially after query rewrite. 
Our \texttt{NOT\_MATERIALIZE} hint directs the optimizer to inline the targeted CTE, expanding its definition into the main query as if it were a FROM-clause subquery. 
This hint enhances query performance by eliminating materialization overhead and enabling more effective plan optimizations, such as improved predicate pushdown. 
Ultimately, these selected and newly introduced hints constitute our query hint base $\mathcal{H}$, detailed in Table~\ref{tab:hintbase}.

\subsection{Hint Injection}
\label{sec:6.2}

This section explains how we use LLMs to identify operations that could benefit from hints and recommend appropriate ones. 
Prior query-level methods~\cite{bao,autosteer,fastgres,steeringqo} apply a single hint to the entire query, but we observe that such a hint can be suboptimal, as a hint that benefits one operation may degrade others. To overcome this limitation, our strategy instead applies hints at the operator level, such as to individual tables or specific operations within a query, to maximize their positive effects. 

Building on the validated rewrite of \(Q'\), we now turn to refining its physical execution plan through targeted hint injection. First, given the physical plan of \(Q'\), database semantics and data distribution statistics,  we identify potential improvements that can be achieved through hints: (1) For every cardinality estimation in the plan, an LLM judges its reasonableness. If an estimation is deemed unreasonable, the LLM provides a recommended range and justification. (2) For all join operations, the LLM assesses whether the selected join operator (such as hash, merge, or nested-loop join) is suitable given the available statistics. 
Second, the Assistant Agent processes these justifications, verifies their consistency against plan statistics, and constructs the final set of hints before forwarding them to the injection function.
Regarding \textit{NOT\_MATERIALIZE}, we apply this hint only if a CTE appears exactly once or processes a small, simple dataset. Finally, the proposed hints are then injected into \(Q'\), producing the final optimized query \(Q^*\).

Unlike general existing methods~\cite{autosteer,bao,steeringqo,fastgres}, our LLM-guided hint injection pinpoints costly operations and generates finer-grained hint recommendations. This enhances the explainability of the recommendations and leads to further improvement. Moreover, in contrast to ML-based hint  approaches~\cite{autosteer,bao,steeringqo,fastgres} that rely on pre-trained models and multi-plan enumeration, our method requires zero training yet shows a competitive performance within a small subset of query hints.
    
\section{Experiment}
\label{sec:7}
In this section, we conduct a comprehensive evaluation of QUITE through performance comparisons, ablation studies, robustness assessments, analyses of rewrite behaviors and cost evaluations.
\subsection{Experimental Set up}
\label{sec:7.1}
\textbf{Testbed.} Our experiments were conducted on a server featuring a 32-core Intel Xeon Platinum 8352V CPU, 251 GB of RAM, a 942 GB SSD, and PostgreSQL v14.13. 

\noindent  \textbf{Workloads.} 
We evaluate QUITE on standard OLAP benchmarks with long-running analytical queries, aligning with our deployment scenarios.
We employ three widely recognized datasets and one LLM-synthesized dataset to evaluate all of the approaches.
 (1) \textbf{TPC-H~\cite{TPC_H_toolkit}} is a well-known OLAP benchmark containing 62 columns and 22 query templates. We generate 63 queries from these templates, excluding Q15 due to its use of CREATE VIEW, a format unsupported by the rewrite engines of two baseline methods~\cite{lr,llm-r2}. We used a scale factor (SF) of 10. (2) \textbf{DSB~\cite{ding2021dsb}} is adapted from the TPC-DS benchmark~\cite{tpcds} and is notable for its complex data distribution and challenging long-context query templates, which consists of 52 query templates. We generate 156 queries based on these templates with $SF=10$. (3) \textbf{Calcite~\cite{apache_calcite_test}} is a real-world workload used to evaluate rewrite rules in Apache Calcite~\cite{apache}. We randomly select 58 queries and generate 10G of data with uniform distributions following previous work~\cite{slabcity}. 
 (4) \textbf{StackOverflow (StackOverflow-Math)~\cite{stackoverflow-math}} is a 13.8 GB dataset from the Mathematics community on the Stack Exchange network\footnote{Stack Overflow is the flagship site of the network~\cite{sqlstorm}}, containing math-related posts and metadata as of October 3, 2024. Following ~\cite{sqlstorm}, we generate queries using predefined complexity prompts (P1–P5) and obtain 43 valid queries after dialect correction.

\noindent  \textbf{Baselines}. We compare our system against the following state-of-the-art query rewrite baselines: 
\begin{itemize}[leftmargin=*, noitemsep, topsep=0pt]
    \item \textbf{LearnedRewrite (LR) (SIGMOD 2022) ~\cite{lr}} uses a Monte Carlo Tree Search (MCTS) algorithm with a learned cost estimation model to explore a SQL policy tree of rewrite rule orders. We adopt the original implementation and use the cost estimation model provided in their official code repository.
    \item \textbf{LLM-R\textsuperscript{2} (VLDB 2025)~\cite{llm-r2}} is a rule-based query rewrite system built on Apache Calcite~\cite{apache}. It uses LLM's In-Context Learning (ICL) ability to select query rewrite rules according to high quality demonstration queries. 
    \item \textbf{R-bot (VLDB 2025)~\cite{r-bot}} is an LLM-based system that leverages an embedded retrieval-augmented generation (RAG) knowledge base and reflective reasoning to select promising rewrite rules.
    \item \textbf{LLM Agent} rewrites queries using an individual LLM agent. To ensure a fair comparison, this LLM agent can access target database statistics and our SQL corrector for self-reflection.
\end{itemize}

\begin{table*}[ht]
    \centering
    \small
    \setlength{\tabcolsep}{3pt} 
    
    \caption{Query Latency of Different Methods (s)}
    \vspace{-1em}
    \label{tab:all}
    \begin{tabular}{lcccccccccccccccc}
        \toprule
        \multicolumn{1}{c}{\multirow{2}{*}{\raisebox{-0.3ex}{\rule{0pt}{2.5ex}Methods}}}
        & \multicolumn{4}{c}{TPC-H (SF=10)}
        & \multicolumn{4}{c}{DSB (SF=10)}
        & \multicolumn{4}{c}{Calcite (SF=10)}
        & \multicolumn{4}{c}{StackOverflow (SF=10)}
        \\
        \cmidrule(lr){2-5} \cmidrule(lr){6-9} \cmidrule(lr){10-13} \cmidrule(lr){14-17}
        & Mean & Median & 75th & 95th 
        & Mean & Median & 75th & 95th
        & Mean & Median & 75th & 95th
        & Mean & Median & 75th & 95th
        \\
        \midrule
        Original   
        & 69.84 & 9.64 & 32.75 & 300.00
        & 32.62 & 4.85 & 10.29 & 300.00
        & 23.88 & 2.58 & 8.64 & 122.36
        &  48.02 & 9.47 & 20.80 & 300.00
        \\
        \midrule

        LearnedRewrite   
        & 37.57 & 9.97 & 30.55 & 202.31
        & 31.93 & 5.14 & 15.18 & 251.42
        & 22.88 & 2.32 & 8.09 & 122.58
        & 46.25 & 9.43 & 18.35 & 300.00 
        \\
        LLM-R\textsuperscript{2}(GPT-4o)
        & 57.09 & 9.30 & 30.50 & 300.00
        & 10.53 & 4.11 & 8.36 & 32.88
        & 16.98 & 2.11 & 7.75 & 50.97
        & 47.08 & 9.19 & 16.06 & 300.00
        \\
        LLM-R\textsuperscript{2}(Claude-3.7)
        & 57.68 & 9.89 & 30.07 & 300.00
        & 9.11 & 3.82 & 8.24 & 27.48
        & 17.80 & 2.30 & 7.83 & 57.28
        & 46.97 & 8.40 & 16.30 & 300.00
        \\
        LLM-R\textsuperscript{2}(DS-R1)
        & 56.73 & 9.25 & 29.33 & 300.00
        & 9.16 & 3.84 & 8.21 & 24.21
        & 16.90 & 2.05 & 7.71 & 49.91
        & 43.42 & 9.21 & 14.14 & 300.00 
        \\
        R-Bot (GPT-4o)
        & 32.64 & 9.48 & 26.25 & 172.20
        & 21.75 & 4.26 & 8.80 & 119.19
        & 22.21 & 2.34 & 8.35 & 122.14
        & 46.43 & 9.37 & 16.02 & 300.00
        \\
        R-Bot (Claude-3.7)
        & 33.89 & 9.71 & 29.32 & 195.51
        & 19.89 & 4.27 & 8.99 & 112.24
        & 23.38 & 2.60 & 8.64 & 122.95
        & 52.06 & 9.52 & 19.25 & 300.00
        \\
        R-Bot (DS-R1)
        & 33.70 & 9.88 & 28.38 & 176.90
        & 21.21 & 4.44 & 9.11 & 113.91
        & 22.71 & 2.52 & 8.43 & 122.89
        & 50.57 & 9.46 & 17.39 & 300.00
        \\
        LLM Agent (GPT-4o)
        & 69.03 & 10.00 & 27.02 & 300.00
        & 33.44 & 4.39 & 9.65 & 300.00
        & 15.47 & 2.03 & 7.32 & 35.34
        & 50.90 & 7.96 & 26.73 & 300.00
        \\
        LLM Agent (DS-R1)
        & 42.79 & 9.94 & 27.95 & 300.00
        & 13.53 & 3.87 & 7.93 & 30.18
        & 15.18 & 1.65 & 7.55 & 36.70
        & 27.14 & 6.38 & 9.40 & 272.55 
        \\
        LLM Agent (DS-V3)
        & 60.12 & \textbf{9.22} & 29.92 & 300.00
        & 32.34 & 4.19 & 9.99 & 300.00
        & 16.49 & 2.53 & 7.72 & 45.15
        & 54.92& 10.33&23.60 &300.00
        \\
        LLM Agent (Claude-3.7)
        & 57.46 & 11.04 & 36.99 & 300.00
        & 25.32 & 4.38 & 8.84 & 300.00
        & 15.67 & 1.59 & 7.68 & 44.54
        & 36.23 & 7.58 & 21.86 & 283.42
        \\
        \midrule
        \system$\circ$ (DS-R1+Claude-3.7)
        & 26.06 & 9.28 & 24.19 & 75.64
        & 6.08 & 3.78 & 7.98 & 21.43
        & 10.00 & \textbf{1.22} & 7.15 & \textbf{26.97}
        & 12.83 & 5.16 & 9.05 & 16.11
        \\
        \system$\star$ (DS-R1+Claude-3.7)
        & \textbf{25.60} & 9.28 & \textbf{23.49} & \textbf{74.43}
        & \textbf{5.85} & \textbf{3.35} & \textbf{7.63} & \textbf{19.97}
        & \textbf{9.97} & \textbf{1.22} & \textbf{6.86} & \textbf{26.97}
        & \textbf{12.61} & \textbf{5.03} & \textbf{7.96} & \textbf{15.89}
        \\
        \bottomrule
    \end{tabular}

    \footnotesize
    \vspace{1em}

    \textbf{Note:} 
    \system$\circ$ means rewriting queries without hint injection \quad
    \system$\star$ means rewriting queries with hint injection \quad
    DS-R1 means DeepSeek-R1

    \vspace{-1em}
\end{table*}

\begin{figure*}[ht]
  \centering
  \begin{minipage}{0.25\textwidth}
    \centering
    \includegraphics[width=\textwidth]{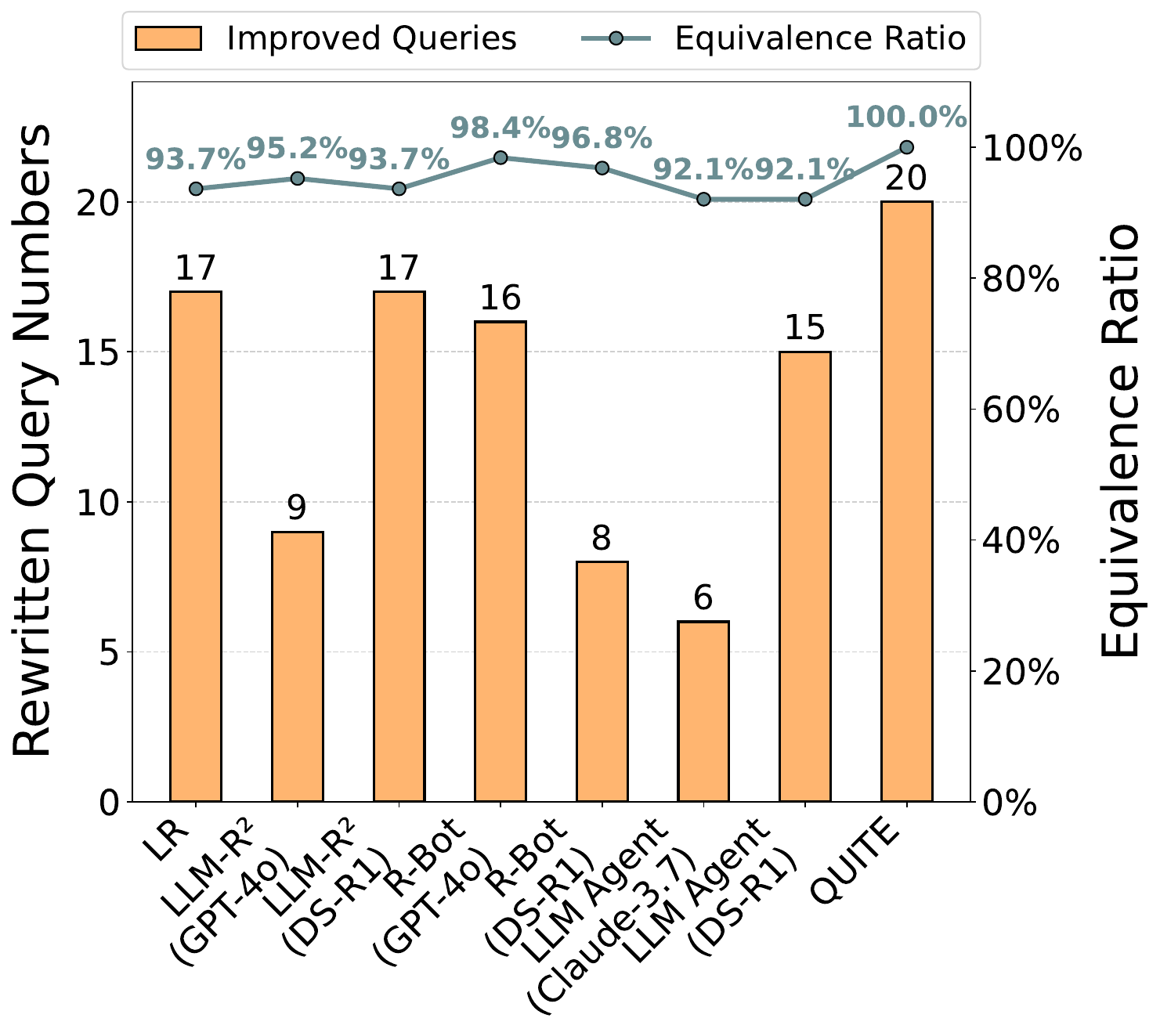}
    \vspace{-2.5em}
    \caption*{(a) TPC-H}
    \label{fig:tpch eqv} 
  \end{minipage}%
  \begin{minipage}{0.25\textwidth}
    \centering
    \includegraphics[width=\textwidth]{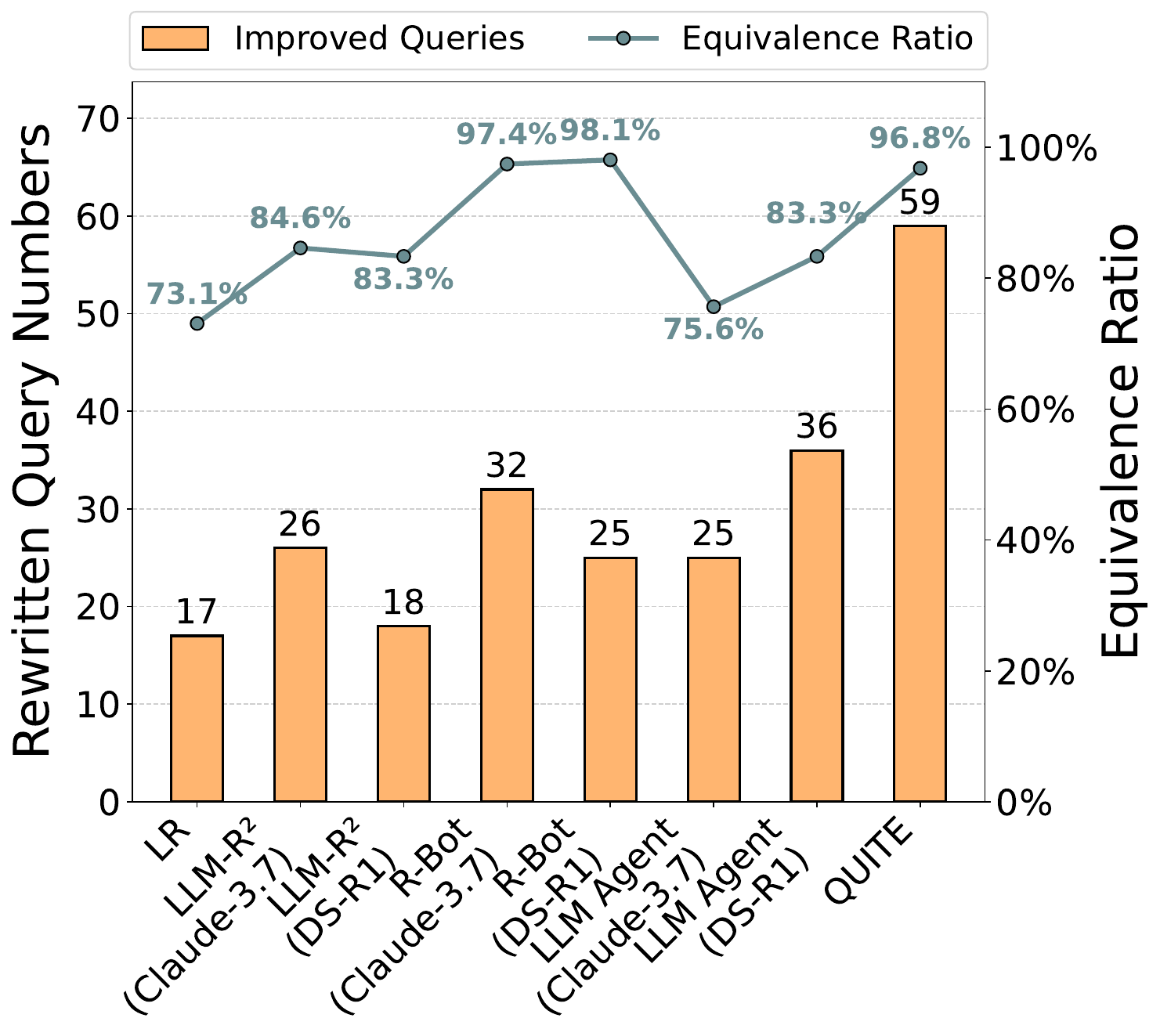}
    \vspace{-2.5em}
    \caption*{(b) DSB}
    \label{fig:dsb} 
  \end{minipage}%
  \begin{minipage}{0.25\textwidth}
    \centering

    \includegraphics[width=\textwidth]{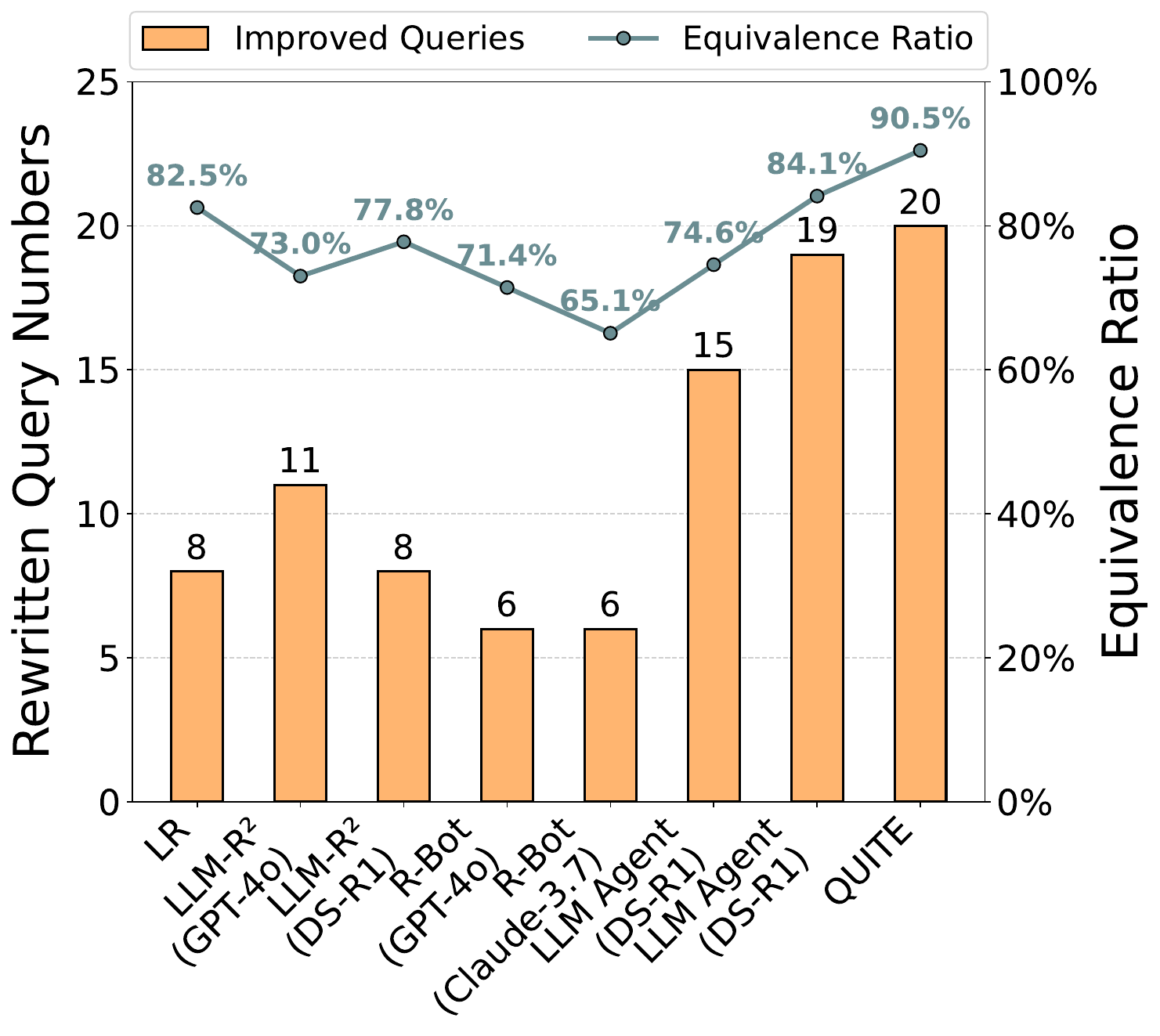}
    \vspace{-2.5em}
    \caption*{(c) Calcite}
    \label{fig:calcite} 
  \end{minipage}%
  \begin{minipage}{0.25\textwidth}
    \centering
    \includegraphics[width=\textwidth]{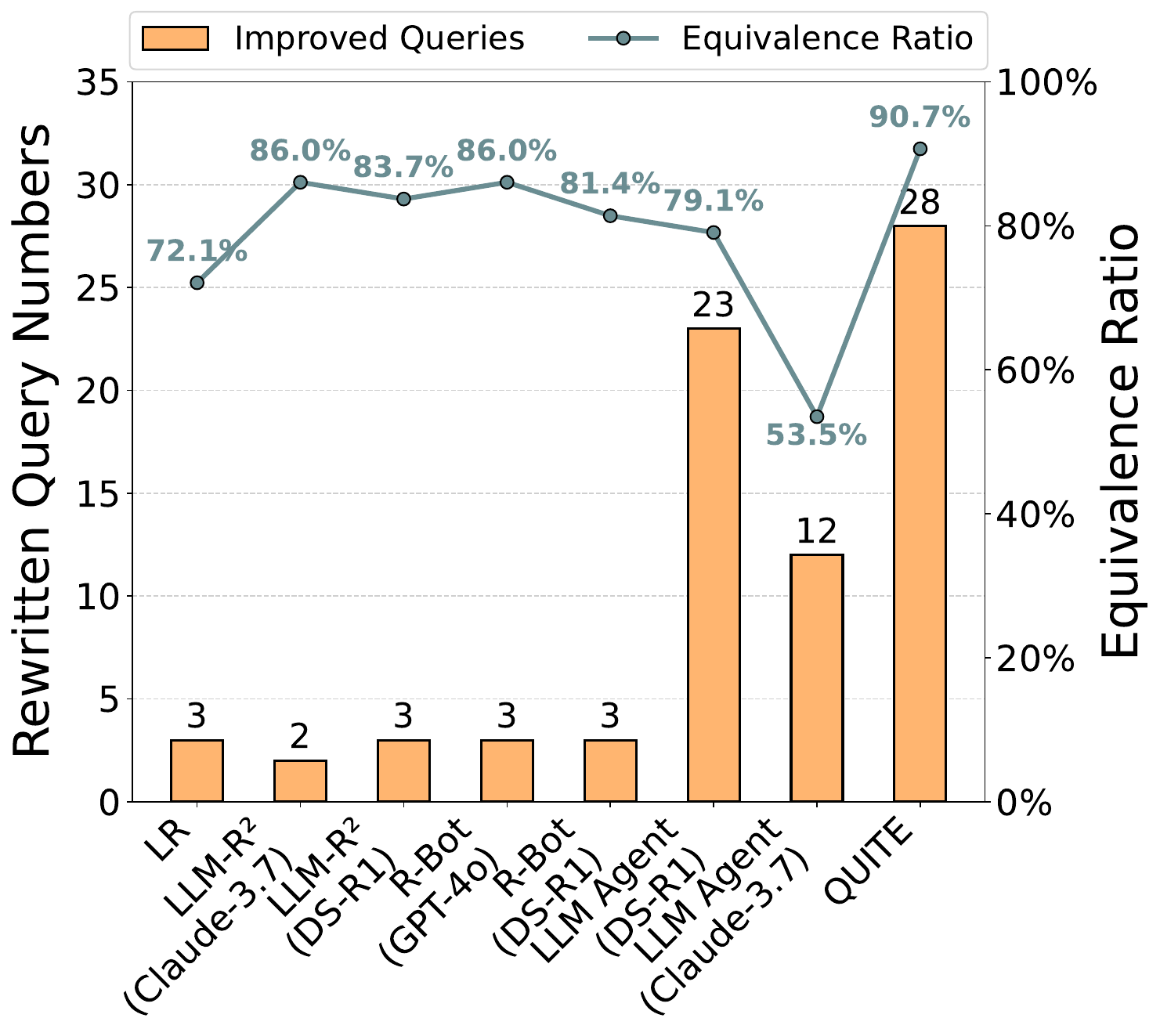}
    \vspace{-2.5em}
    \caption*{(d) StackOverflow}
    \label{fig:sqlstorm}
  \end{minipage}
  \vspace{-1em}
  \caption{Rewrite Equivalence and Improvement Numbers on Different Benchmarks} 
  \label{fig:benchmarks} 
  \vspace{-1em}
\end{figure*} 

\begin{figure}[t]
    \centering
    \includegraphics[width=1.0\linewidth]{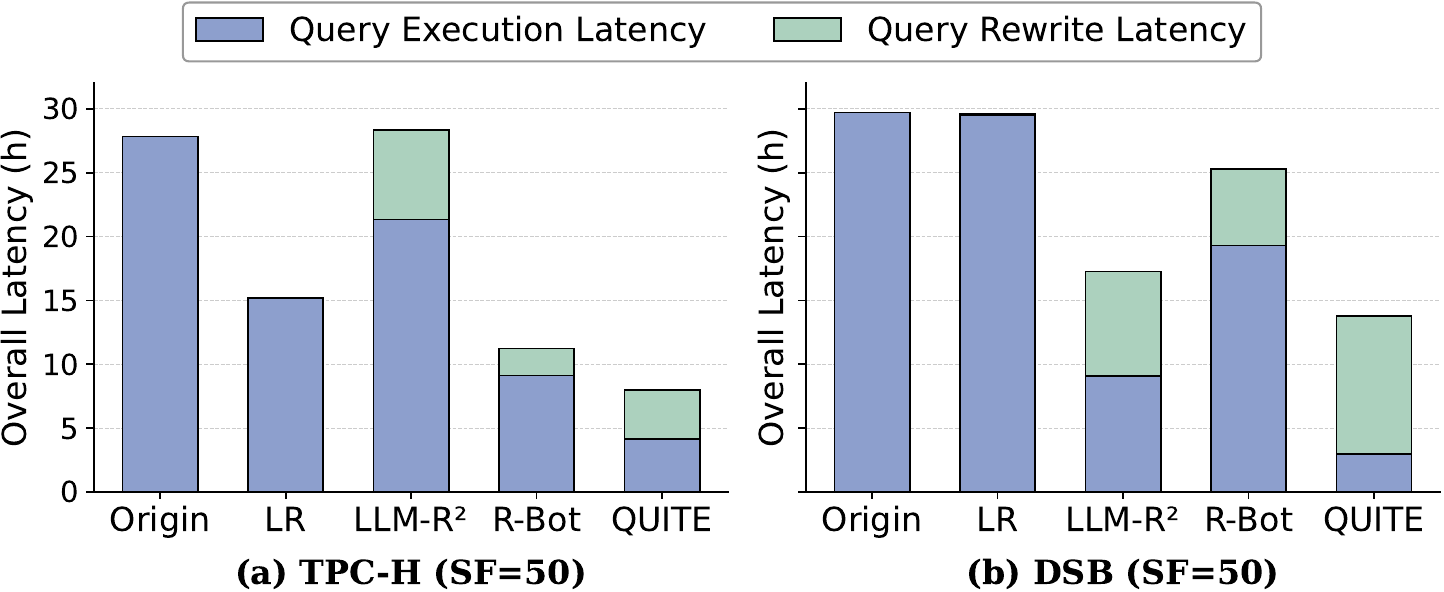}
        \vspace{-2em}
    \caption{Comparison of Overall Latency (h)}
    \label{fig:dataset_50x}
    \vspace{-2em}
\end{figure}

\noindent \textbf{Comparison with GenRewrite~\cite{genrewrite}}. 
GenRewrite explores query rewriting with natural-language rewrite rules and iterative correction, but the lack of released code prevents direct comparison. 
At a high level, it accumulates natural-language rewrite knowledge to guide subsequent LLM-based rewriting.
In contrast, QUITE formulates query rewrite as a decision process and directly optimizes execution through iterative refinement with database feedback, enabling more robust and data-aware optimizations.

\noindent  \textbf{\system\ Settings.} We use DeepSeek-R1 as the reasoning agent due to its strong analysis capability, and Claude-3.7-Sonnet as the rewrite, decision, and assistant agents by default. A temperature of 0 is applied to all LLMs to ensure output stability~\cite{DeepSeek_doc,anthropic_doc}.
The number of iterations is limited to 2 to prevent excessive FSM rewrite cycles. All LLM-involved experiments employ the identical temperature configurations used in our approach.

\noindent  \textbf{Evaluation Metrics.} 
(1) \textit{Query Execution Latency.} The duration required to complete a query, reported using the average, median, and 95th percentile latency values.
(2) \textit{Rewritten Equivalence Rate.} The fraction of rewritten queries that produce results matching those of the original queries. In our work, equivalence is determined by comparing the execution outputs of the original and rewritten queries.
(3) \textit{Rewritten Improvement Rate.} The percentage of rewritten SQL queries that show a performance improvement. A performance gain is considered significant if it results in at least a 10\% reduction in execution time.

\noindent  \textbf{Query Evaluation Approach}. 
Before each query, the database is restarted to clear caches. The query execution process involves an initial warm-up run to mitigate cold start impacts, followed by three measured executions. We report the average execution time of these three runs. Queries exceeding 300 seconds are terminated and recorded~\cite{llm-r2,r-bot}, with capped values included in all latency statistics (mean, median, 75th, and 95th percentiles).
Rewritten queries with syntax errors or output mismatches are marked as non-equivalent. In such cases, the execution time of the original query is used.
If a rewrite times out, we recheck equivalence on a smaller scale: non-equivalent rewrites fall back to the original, while equivalent rewrites are executed with capped latency.
\vspace{-1em}
\subsection{Performance Comparison}
\label{sec:7.2}
\noindent \textbf{Query Execution Latency.}
As shown in Table \ref{tab:all}, queries optimized by \system\ demonstrate significantly reduced execution times, consistently outperforming all baseline methods. Across the TPC-H, DSB, Calcite and StackOverflow benchmarks, \system\ achieves substantial average execution time reductions. Specifically, \system\ reduces execution time by 31.9\%, 81.7\%, 56.4\% and 72.7\% compared to LR; 54.9\%, 35.8\%, 41.0\% and 71.0\% over LLM-R\textsuperscript{2}; and 40.2\%, 56.8\%, 34.3\% and 53.5\% against LLM Agent. When evaluated against R-Bot, \system\ achieves reductions of 21.6\%, 70.6\%, 55.1\% and and 72.8\%. \system's advantages are further illustrated by the rewritten improvement rates shown in Figure \ref{fig:benchmarks}, where \system\ achieves the highest ratios of 31.7\%, 37.8\%, 34.5\% and 65\% on TPC-H, DSB, and Calcite, respectively, indicating strong generalization capabilities across different workloads.

The optimization benefits of QUITE are more pronounced on complex datasets. Notably, on the LLM-generated StackOverflow workload, we did not curate queries to disadvantage baselines. Since many LLM-generated queries use broad SQL dialects that are not directly supported by the strict syntax of Apache Calcite used in rule-based pipelines~\cite{lr,llm-r2,r-bot}, we minimally normalized only such dialect mismatches (e.g., \texttt{LISTAGG} vs. \texttt{STRING\_AGG}), discarding unresolvable cases. Under this controlled setup, rule-based pipelines remain bottlenecked by dialect constraints and limited rule and pattern coverage on complex structures, while QUITE is more robust and delivers larger latency gains.



\begin{table*}[htbp]
\centering
\caption{Unified Ablation Study Results on DSB (SF=10)}
\vspace{-0.8em}
\footnotesize
\label{tab:unified-ab}
\renewcommand{\arraystretch}{1.12}
\setlength{\tabcolsep}{3pt}
\begin{tabular}{lcccccccccccccc}
\toprule
\multirow{2}{*}{\textbf{Metrics}} &
\multicolumn{3}{c}{\textbf{Structured Knowledge Base}} &
\multicolumn{5}{c}{\textbf{Different LLMs}} &
\multicolumn{5}{c}{\textbf{Hint Construction Methods}} \\
\cmidrule(lr){2-4} \cmidrule(lr){5-9} \cmidrule(lr){10-14}
 & \makecell{\textbf{w/o}\\\textbf{Q\&A Units}} 
 & \makecell{\textbf{Raw}\\\textbf{Q\&A Units}} 
 & \makecell{\textbf{Filtered}\\\textbf{Q\&A Units}}
 & \makecell{\textbf{DS-R1 +}\\\textbf{GPT-4o}} 
 & \makecell{\textbf{DS-R1 +}\\\textbf{DS-V3}} 
 & \makecell{\textbf{DS-R1 +}\\\textbf{Claude-3.7}} 
 & \makecell{\textbf{DS-V3 +}\\\textbf{Claude-3.7}} 
 & \makecell{\textbf{Claude-3.7 +}\\\textbf{Claude-3.7}} 
 & \makecell{\textbf{w/o}\\\textbf{Hints}} 
 & \makecell{\textbf{Bao’s}\\\textbf{Hint Base}} 
 & \makecell{\textbf{Full Hints}\\\textbf{(no GUC)}} 
 & \makecell{\textbf{Ours}\\\textbf{+ Scan}}
 & \makecell{\textbf{Our}\\\textbf{Hint Base}} 
 \\ 
\midrule
Mean   & 9.96 & 9.10 & \textbf{6.08} & 31.38 & 28.07 & \textbf{6.08} & 23.41 & 12.76 & 6.08 & 6.93  & 7.06 & 6.04 & \textbf{5.85} \\
Median & 3.94 & 3.85 & \textbf{3.78} & 4.64  & 4.43  & \textbf{3.78} & 4.32 & 4.01 & 3.78 & 4.65  & 4.02 & 3.89 & \textbf{3.35} \\
75th   & 8.49 & 8.10 & \textbf{7.98} & 9.56  & 9.07  & \textbf{7.98} & 9.64 & 8.74 & 7.98 & 8.46  & 8.64 & 8.06  & \textbf{7.63} \\
95th   & 24.81 & 23.01 & \textbf{21.43} & 300.00 & 300.00 & \textbf{21.43} & 148.47 & 58.38 & 21.43 & 20.41  & 22.52 & 20.29 & \textbf{19.97} \\
\bottomrule
\end{tabular}
\vspace{-0.8em}
\end{table*}

\begin{figure*}[!t]
  \centering
  \includegraphics[width=1\textwidth]{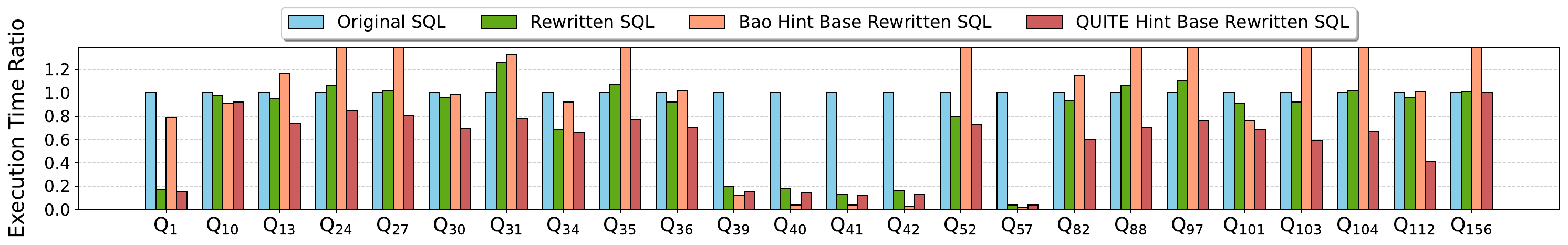} 
  \vspace{-2em}
  \caption{Ablation Study of Hint Injection on DSB Benchmark}
  \label{fig:hint aware}
  \vspace{-0.5em}
\end{figure*}

\begin{figure}[t]
    \centering
    \includegraphics[width=1.0\linewidth]{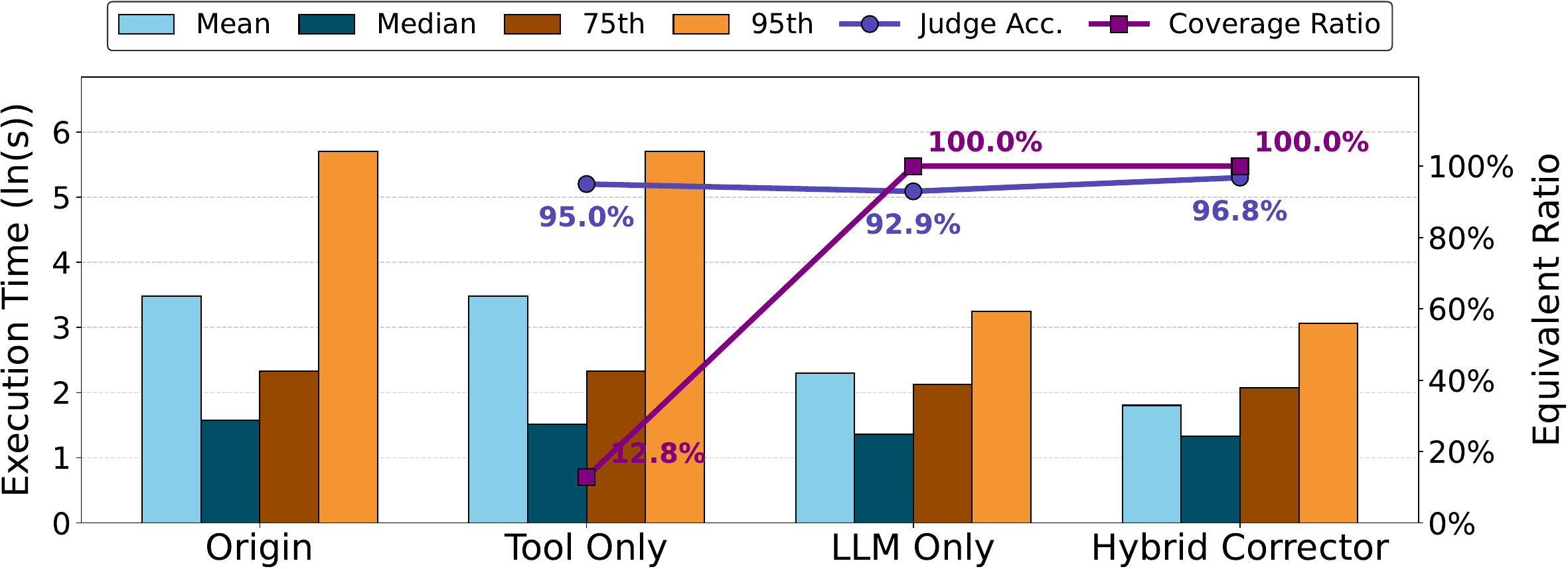}
        \footnotesize

            \vspace{-1em}
    \caption{Ablation Study of Hybrid SQL
Corrector}
    \label{fig:tools_ablation}
    \vspace{-2.5em}
\end{figure}

\begin{table}[t]
  \centering
  \caption{Ablation Study of the Reasoning Agent and FSM}
  \vspace{-1em}
  \label{tab:ablation_FSM}
  \begin{tabular}{@{}ccrrrr@{}}
    \toprule
    \multicolumn{2}{c}{Methods} 
      & \multicolumn{4}{c}{DSB (SF=10)} \\
    \cmidrule(lr){1-2} \cmidrule(lr){3-6}
    \makecell{MDP-based \\ Reasoning Agent} & FSM 
      & Mean & Median & 75th & 95th \\
    \midrule
    \xmark & \xmark & 13.48             & 4.84            &  10.22            & 33.64             \\
    \xmark & \cmark & 12.72          & 4.80           & 9.93          & 29.91          \\
    \cmark & \cmark & \textbf{6.08}  & \textbf{3.78}  & \textbf{7.98}  & \textbf{21.43} \\
    \bottomrule
  \end{tabular}
  \vspace{-3em}
\end{table}

\noindent \textbf{Overall Latency}. 
As shown in Figure~\ref{fig:dataset_50x}, we evaluate overall execution time on TPC-H (SF=50) and DSB (SF=50) with a per-query time limit of 8000 seconds, using Claude-3.7-Sonnet across all baselines. On TPC-H and DSB, \system\ reduced execution time by 71.4\% and 53.6\% compared with the original queries, outperforming LR (47.6\%, 53.4\%), LLM-R\textsuperscript{2} (71.9\%, 20.0\%), and R-Bot (29.1\%, 45.5\%), respectively. 
These results demonstrate \system’s strong end-to-end advantage. \system’s rewrite time on the DSB dataset is relatively longer, which can be attributed to two main factors: first, DeepSeek-R1 requires more time to process difficult problems; second, for complex queries, the decision agent often performs an additional iteration to explore further optimizations.



\noindent \textbf{Rewritten Equivalence.} Figure \ref{fig:benchmarks} details the number of equivalent and improved rewrites produced by each method across the datasets. Notably, direct LLM-agent rewrites, especially by Claude-3.7-Sonnet on StackOverflow, often violate equivalence due to subtle semantic drift during query restructuring. In contrast, \system\ achieves substantially higher equivalence rates, attaining 100\%, 96.8\%, 98.3\% and 90.7\% on TPC-H, DSB, Calcite and StackOverflow. In addition, rule-based methods suffer from incomplete semantic equivalence, as many rules in their rewrite engines (e.g., Apache Calcite~\cite{apache}) lack formal verification due to the inherent difficulty of equivalence checking~\cite{wetune,sqlsolver}. Despite this, our approach achieves the highest equivalence rate across all benchmarks, advancing the state of the art.


\subsection{Ablation Study}
\label{sec:7.3}
We design our ablation study from four perspectives by selectively replacing or removing system components.

\noindent \textbf{Rewrite Middleware.} First, we ablated the hybrid SQL corrector to measure its effect on equivalence and execution. Figure~\ref{fig:tools_ablation} reports Judge Accuracy (the fraction of correct decisions) and Coverage Ratio (the fraction of queries with a definitive decision). Its removal significantly degrades performance: nonequivalent queries increase from 5 to 11, unchanged queries increase by 13, and improved queries drop by 17. These results highlight the corrector's essential role in ensuring query quality and efficiency.

We also evaluate the impact of the structured knowledge base. As shown in Table~\ref{tab:unified-ab}, \system\ consistently outperforms its variant without this module across all metrics. The improvement stems from high-quality domain knowledge embedded in the knowledge base, which effectively guide more optimized rewrites. 
To justify the LLM-based filtering, we compare the 241 filtered units against all 3,432 collected units. The filtered version delivers significantly stronger execution performance: five queries achieve a $2\times$–$5\times$ speedup, and more than eight queries exceed $5\times$. This confirms that consensus-aligned, high-quality knowledge provides significantly more effective guidance for LLM rewriting, enabling stable and efficient query optimization.

\begin{table*}[t]
\centering
\caption{Robustness Study on TPC-H and DSB Benchmarks under Different Scale Factors (SF)}
\vspace{-0.6em}
\label{tab:robustness_all}
\scriptsize
\setlength{\tabcolsep}{2.6pt} 
\begin{tabular}{lcccccccccccccccccccccccc}
\toprule
\multirow{3}{*}{Methods} &
\multicolumn{12}{c}{\textbf{TPC-H}} &
\multicolumn{12}{c}{\textbf{DSB}} \\
\cmidrule(lr){2-13} \cmidrule(lr){14-25}
 & \multicolumn{4}{c}{SF=1} & \multicolumn{4}{c}{SF=10} & \multicolumn{4}{c}{SF=30} 
 & \multicolumn{4}{c}{SF=1} & \multicolumn{4}{c}{SF=10} & \multicolumn{4}{c}{SF=30} \\
\cmidrule(lr){2-5} \cmidrule(lr){6-9} \cmidrule(lr){10-13} 
\cmidrule(lr){14-17} \cmidrule(lr){18-21} \cmidrule(lr){22-25}
 & Mean & Med & 75th & 95th & Mean & Med & 75th & 95th & Mean & Med & 75th & 95th
 & Mean & Med & 75th & 95th & Mean & Med & 75th & 95th & Mean & Med & 75th & 95th \\
\midrule
Original & 29.48 & 0.69 & 1.20 & 300.00 & 69.31 & 9.42 & 30.08 & 300.00 & 64.76 & 20.61 & 46.69 & 300.00
& 7.22 & 0.27 & 0.69 & 13.19 & 32.62 & 4.85 & 10.29 & 300.00 & 39.42 & 5.06 & 18.22 & 300.00 \\
\midrule
LearnedRewrite & 1.46 & 0.52 & 1.10 & 3.91 & 37.57 & 9.97 & 30.55 & 202.31 & 52.59 & 18.90 & 43.50 & 300.00
& 17.15 & 0.42 & 1.52 & 81.14 & 31.93 & 5.14 & 15.18 & 251.42 & 43.88 & 7.09 & 20.73 & 300.00 \\
LLM-R\textsuperscript{2}(DS-R1) & 15.38 & 0.69 & 1.22 & 7.69 & 56.73 & \textbf{9.25} & 29.33 & 300.00 & 54.67 & 18.30 & 58.11 & 300.00
& 4.30 & \textbf{0.23} & 0.57 & 2.15 & 9.16 & 3.84 & 8.21 & 24.21 & 23.01 & 5.01 & 11.87 & 87.15 \\
R-Bot (GPT-4o) & 1.49 & 0.66 & 1.03 & 6.38 & 32.64 & 9.48 & 26.25 & 172.20 & 54.54 & 19.85 & 61.12 & 300.00
& 9.06 & 0.25 & 0.63 & 21.89 & 21.75 & 4.26 & 8.80 & 119.19 & 31.93 & 5.15 & 16.27 & 300.00 \\
\midrule
\system\ (Rw. on SF=1) & \textbf{1.02} & 0.64 & \textbf{1.01} & \textbf{3.79} & 36.87 & 13.53 & 29.02 & 279.51 & 54.36 & 25.02 & 43.03 & 289.48
& \textbf{2.25} & 0.24 & \textbf{0.50} & 2.74 & 10.58 & 4.18 & 8.59 & 30.79 & 23.23 & 4.67 & 12.91 & 111.83 \\
\system\ (Rw. on SF=10) & 1.08 & 0.68 & 1.22 & \textbf{3.79} & \textbf{26.06} & 9.28 & \textbf{24.19} & \textbf{75.64} & 45.90 & 17.44 & 57.26 & 155.72
& 2.82 & 0.24 & 0.51 & \textbf{1.72} & \textbf{6.08} & \textbf{3.78} & \textbf{7.98} & \textbf{21.43} & 17.36 & \textbf{4.34} & 9.28 & 77.62 \\
\system\ (Rw. on SF=30) & 5.72 & \textbf{0.50} & 1.17 & 5.17 & 30.50 & 9.96 & 24.29 & 96.68 & \textbf{40.65} & \textbf{16.89} & \textbf{37.96} & \textbf{131.97}
& 6.65 & 0.26 & 0.56 & 3.20 & 8.74 & 3.94 & 8.01 & 23.15 & \textbf{16.35} & 4.45 & \textbf{9.26} & \textbf{74.33} \\
\bottomrule
\end{tabular}
\vspace{-1em}
\end{table*}

\noindent \textbf{LLM-Agent-based Query Rewrite FSM.}
To evaluate the impact of the FSM framework, we make two modifications: (1) remove the equivalence-checking loop and (2) remove the decision stage. This converts the process into a linear pipeline while keeping all components and agent tools for rewriting. To further isolate the contribution of the MDP-based reasoning agent, we also conduct an ablation that removes the MDP component. As shown in Table \ref{tab:ablation_FSM}, \system\ outperforms all variants across metrics. Notably, adding the MDP reduces the average execution time by 52.2\% compared to the version without it, indicating that the MDP substantially improves the reasoning agent’s effectiveness.

\noindent \textbf{Query Hint Injection.}
We evaluate the impact of hint injection on queries rewritten by \system. As shown in Table \ref{tab:all}, hint injection further reduces average execution time by 1.8\%, 3.8\%, and 0.3\% on TPC-H, DSB, and Calcite, respectively, indicating that appropriate hints enable additional fine-grained optimization even after extensive rewriting. To highlight the advantage of our approach, we compare it with using LLMs to select Bao’s GUC-level hint sets. As shown in Table \ref{tab:unified-ab} and Figure \ref{fig:hint aware}, this baseline exhibits an overall performance degradation within Bao’s predefined hint space. Furthermore, Figure \ref{fig:hint aware} shows execution times for 24 positively optimized DSB queries before and after hint injection, as well as for the original queries. In several cases, hint injection recovers and surpasses the original performance when rewriting alone is suboptimal. This demonstrates hint injection can recover and enhance performance in cases where the initial rewrites were suboptimal.

\begin{figure}[t]
    \centering
    \includegraphics[width=1.0\linewidth]{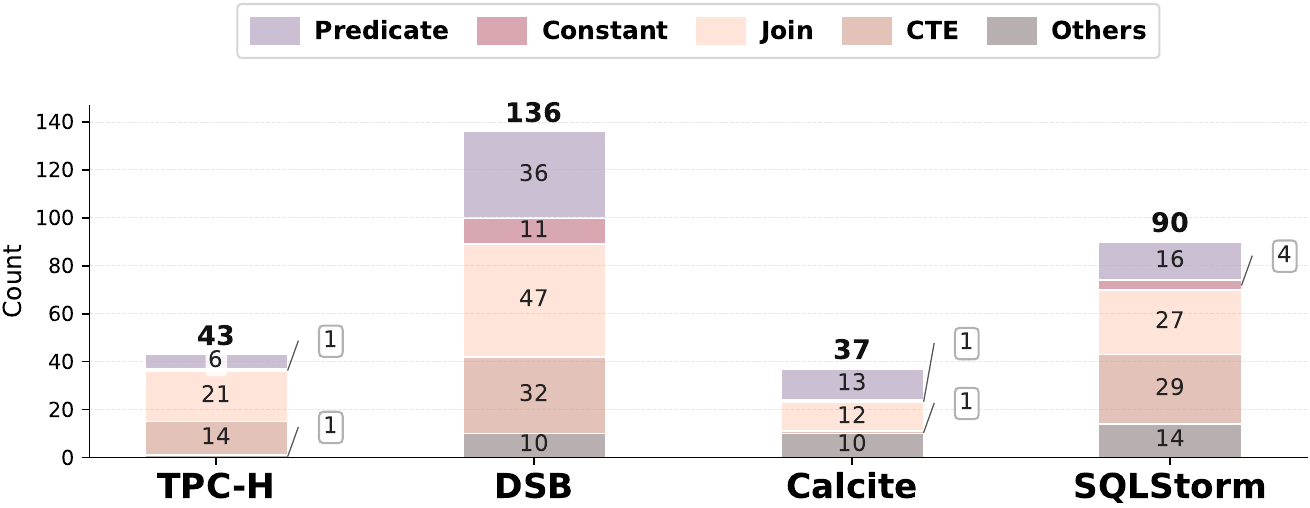}
    \footnotesize
    \vspace{-1.0em}
    \caption{Rewrite Categories for Different Benchmarks}
    \label{fig:rewrite_s}
    \vspace{-1em}
\end{figure}

\begin{table}[t]
\centering
\caption{Cost Analysis on TPC-H Benchmark}
\vspace{-1em}

\label{tab:time_and_cost}
\begin{tabular}{lcc}
\toprule
Methods & Average Time (s) & Cost (\$) \\ 
\midrule

LLM-R\textsuperscript{2} (Claude-3.7)   & 391.23                     &    0.016        \\ 
R-Bot (Claude-3.7)             & 118.33        &     0.417
\\
 \system\ (DS-R1 + Claude-3.7)          & 218.25                        & 0.200                 \\ 
\bottomrule
\end{tabular}
\vspace{-1em}
\end{table}

\noindent \textbf{Sensitivity Study.}
We examine \system’s sensitivity to the number of rewrite rounds, hint thresholds, and different LLM models. We first test different model combinations. We fix DeepSeek-R1 as the reasoning model and vary the other models among Claude-3.7-Sonnet, DeepSeek-V3, and GPT-4o, and then we replace the reasoning model with DeepSeek-V3 and Claude-3.7-Sonnet. As shown in Table~\ref{tab:unified-ab}, the combination of DeepSeek-R1’s strong reasoning capabilities and Claude-3.7-Sonnet’s long-context understanding significantly outperforms all other combinations.


For hint thresholds, as shown in Table~\ref{tab:unified-ab}, we test the effect of adding the scan method to our hint base and the effect of using all hints (excluding GUC). The former yields an execution time nearly identical to that of the rewritten query, while the latter causes a slight performance degradation. This suggests that the size of our current hint search space is appropriate.



\subsection{Robustness Study}
\label{sec:7.4}
\noindent \textbf{Scale-Aware Robustness.} We evaluate \system\ on TPC-H and DSB under scale factors $SF\in\{1,10,30\}$, where rewriting and execution are performed on the same scale. As Table~\ref{tab:robustness_all} shows, \system\ consistently achieves substantial latency reductions at every scale. In contrast, LR, LLM-R\textsuperscript{2}, and R‑Bot often degrade as the data scale changes.  These results demonstrate that \system\ maintains stable efficiency under varying data volumes and adapts robustly to distribution shifts within each benchmark.

\noindent \textbf{Cross-Scale Robustness.} 
 To examine generalization across data distributions, we perform cross-scale evaluation: rewrites produced at one scale ( $SF\in\{1,10,30\}$) are executed on all others. As Table~\ref{tab:robustness_all} shows, \system\ achieves its strongest gains when rewrite and execution scales match, while mismatched cases show only minor regressions and still outperform all baselines. 
This behavior highlights \system’s strong data adaptivity and robustness, effectively learning to tailor rewrites to underlying data characteristics without overfitting to a specific workload or scale.

\subsection{Further Analysis}
\label{sec:7.6}

\noindent \textbf{Rewrite Type Analysis.} We report successful rewrites that achieve over a 10\% performance improvement while maintaining equivalence across four benchmarks. As shown in Figure~\ref{fig:rewrite_s}, most gains come from join, CTE, and predicate rewrites, while the rest are benchmark-specific. Compared with rule-based baselines~\cite{lr,llm-r2,r-bot}, \system\ produces deeper rewrites beyond pattern matching. We further analyse efficient rewrite types, query structures, Q\&A units, and query hints in our Github files~\cite{quite_ap,quite_rt}.

 \noindent \textbf{Cost Analysis.} 
 We conducted a cost analysis experiment on 63 queries of TPC-H. As shown in Table \ref{tab:time_and_cost}, LLM-R\textsuperscript{2} incurs high time cost for demonstration construction, while R-Bot consumes substantial tokens for stepwise rewriting. In contrast, \system\ offers a balanced trade-off between time expenditure and cost. By generating much of its output with DeepSeek-R1, \system\ greatly reduces monetary cost. Additionally, it avoids pre-training and extensive knowledge base construction. 

\section{Conclusion}
\label{sec:8}
This paper presents \system, a training-free and feedback-aware system that smartly leverages LLM Agents to rewrite SQL queries into semantically equivalent forms with significantly improved performance. We rethink the query rewrite process and identify unique opportunities to support a wider range of query patterns and rewrite strategies with strong scalability. 
The implemented \system~ system significantly outperforms state-of-the-art methods in both query performance and query coverage on widely used benchmarks and synthetic workloads.

\bibliographystyle{ACM-Reference-Format}
\bibliography{sample}

@String{Computing = "Computing" }

@String{Computer = "{IEEE} Computer" }

@article{sqlsolver,
  title={Proving query equivalence using linear integer arithmetic},
  author={Ding, Haoran and Wang, Zhaoguo and Yang, Yicun and Zhang, Dexin and Xu, Zhenglin and Chen, Haibo and Piskac, Ruzica and Li, Jinyang},
  journal={Proceedings of the ACM on Management of Data},
  volume={1},
  number={4},
  pages={1--26},
  year={2023},
  publisher={ACM New York, NY, USA}
}

@article{hottsql,
  title={HoTTSQL: Proving query rewrites with univalent SQL semantics},
  author={Chu, Shumo and Weitz, Konstantin and Cheung, Alvin and Suciu, Dan},
  journal={Acm sigplan notices},
  volume={52},
  number={6},
  pages={510--524},
  year={2017},
  publisher={ACM New York, NY, USA}
}

@article{extensible,
  title={Extensible/rule based query rewrite optimization in Starburst},
  author={Pirahesh, Hamid and Hellerstein, Joseph M and Hasan, Waqar},
  journal={ACM Sigmod Record},
  volume={21},
  number={2},
  pages={39--48},
  year={1992},
  publisher={ACM New York, NY, USA}
}

@article{deepseek-r1,
  title={Deepseek-r1: Incentivizing reasoning capability in llms via reinforcement learning},
  author={Guo, Daya and Yang, Dejian and Zhang, Haowei and Song, Junxiao and Zhang, Ruoyu and Xu, Runxin and Zhu, Qihao and Ma, Shirong and Wang, Peiyi and Bi, Xiao and others},
  journal={arXiv preprint arXiv:2501.12948},
  year={2025}
}

@article{neo,
  title={Neo: A learned query optimizer},
  author={Marcus, Ryan and Negi, Parimarjan and Mao, Hongzi and Zhang, Chi and Alizadeh, Mohammad and Kraska, Tim and Papaemmanouil, Olga and Tatbul, Nesime},
  journal={arXiv preprint arXiv:1904.03711},
  year={2019}
}

@article{database,
  title={Database meets artificial intelligence: A survey},
  author={Zhou, Xuanhe and Chai, Chengliang and Li, Guoliang and Sun, Ji},
  journal={IEEE Transactions on Knowledge and Data Engineering},
  volume={34},
  number={3},
  pages={1096--1116},
  year={2020},
  publisher={IEEE}
}

@inproceedings{bao,
  title={Bao: Making learned query optimization practical},
  author={Marcus, Ryan and Negi, Parimarjan and Mao, Hongzi and Tatbul, Nesime and Alizadeh, Mohammad and Kraska, Tim},
  booktitle={Proceedings of the 2021 International Conference on Management of Data},
  pages={1275--1288},
  year={2021}
}

@inproceedings{query-rewrite,
  title={A rule-based query rewriter in an extensible dbms},
  author={Finance, Beatrice and Gardarin, Georges},
  booktitle={Proceedings. Seventh International Conference on Data Engineering},
  pages={248--249},
  year={1991},
  organization={IEEE Computer Society}
}

@misc{pg_hint_plan,
  author = {{NTT OSS Center DBMS Development and Support Team}},
  title = {{pg\_hint\_plan: Extension adding support for optimizer hints in PostgreSQL}},
  year = {2012},
  url = {https://github.com/ossc-db/pg_hint_plan},
  note = {Accessed: 2025-05-01}
}

@misc{mysql_hints,
  author = {{Oracle Corporation}},
  title = {{MySQL 8.0 Optimizer Hints}},
  year = {2024},
  url = {https://dev.mysql.com/doc/refman/8.0/en/optimizer-hints.html},
  note = {Accessed: 2025-05-29}
}

@misc{oracle_hints,
  author = {{Oracle Corporation}},
  title = {{Oracle Database 19c Optimizer Hints}},
  year = {2019},
  url = {https://docs.oracle.com/en/database/oracle/oracle-database/19/tgsql/optimizer-hints.html},
  note = {Accessed: 2025-05-29}
}

@article{Leis2015,
  author    = {V. Leis and A. Kemper and T. Neumann and P. Boncz and S. Zdonik},
  title     = {How Good Are Query Optimizers, Really?},
  journal   = {Proceedings of the VLDB Endowment},
  volume    = {8},
  number    = {12},
  pages     = {1850--1861},
  year      = {2015},
  url       = {http://vldb.org/pvldb/vol8/p1850-leis.pdf}
}

@misc{quite_rt,
  author = {{Yuyang Song}},
  title = {{QUITE: Rewrite Type Analysis and Taxonomy}},
  year = {2025},
  url = {https://github.com/Yuyang-Song/QUITE/blob/main/documents/rewrite_types.md},
}

@article{codes,
  title={Codes: Towards building open-source language models for text-to-sql},
  author={Li, Haoyang and Zhang, Jing and Liu, Hanbing and Fan, Ju and Zhang, Xiaokang and Zhu, Jun and Wei, Renjie and Pan, Hongyan and Li, Cuiping and Chen, Hong},
  journal={Proceedings of the ACM on Management of Data},
  volume={2},
  number={3},
  pages={1--28},
  year={2024},
  publisher={ACM New York, NY, USA}
}

@article{xiyansql,
  title={Xiyan-sql: A multi-generator ensemble framework for text-to-sql},
  author={Gao, Yingqi and Liu, Yifu and Li, Xiaoxia and Shi, Xiaorong and Zhu, Yin and Wang, Yiming and Li, Shiqi and Li, Wei and Hong, Yuntao and Luo, Zhiling and others},
  journal={arXiv e-prints},
  pages={arXiv--2411},
  year={2024}
}

@article{omnisql,
  title={Omnisql: Synthesizing high-quality text-to-sql data at scale},
  author={Li, Haoyang and Wu, Shang and Zhang, Xiaokang and Huang, Xinmei and Zhang, Jing and Jiang, Fuxin and Wang, Shuai and Zhang, Tieying and Chen, Jianjun and Shi, Rui and others},
  journal={arXiv preprint arXiv:2503.02240},
  year={2025}
}

@misc{quite_ap,
  author = {{Yuyang Song}},
  title = {{QUITE: Appendix}},
  year = {2025},
  url = {https://github.com/Yuyang-Song/QUITE/blob/main/documents/QUITE_Appendix.pdf},
}

@article{base,
  title={Base: Bridging the gap between cost and latency for query optimization},
  author={Chen, Xu and Wang, Zhen and Liu, Shuncheng and Li, Yaliang and Zeng, Kai and Ding, Bolin and Zhou, Jingren and Su, Han and Zheng, Kai},
  journal={Proceedings of the VLDB Endowment},
  volume={16},
  number={8},
  pages={1958--1966},
  year={2023},
  publisher={VLDB Endowment}
}

@inproceedings{cosette,
  title={Cosette: An Automated Prover for SQL.},
  author={Chu, Shumo and Wang, Chenglong and Weitz, Konstantin and Cheung, Alvin},
  booktitle={CIDR},
  pages={1--7},
  year={2017}
}

@article{qed,
  title={QED: A Powerful Query Equivalence Decider for SQL},
  author={Wang, Shuxian and Pan, Sicheng and Cheung, Alvin},
  journal={Proceedings of the VLDB Endowment},
  volume={17},
  number={11},
  pages={3602--3614},
  year={2024},
  publisher={VLDB Endowment}
}

@misc{TPC_H_toolkit,
  title = {TPC-H Toolkit},
  year = {[n.d.]},
  url = {https://www.tpc.org/tpc_documents_current_versions/current_specifications5.asp},
  note = {Accessed: 2024-11-12}
}

@misc{apache_calcite_test,
  title = {Apache Calcite},
  year = {2021},
  url = {https://github.com/georgia-tech-db/spes/blob/main/testData/calcite_tests.json},
  note = {Accessed: 2025-1-22}
}

@misc{stackoverflow-math,
  title = {Stack Exchange inc.2024.},
  url = {https://math.stackexchange.com/},
  note = {Accessed: 2025-10-07}
}

@misc{stackoverflow,
  title        = {Stack Overflow},
  year         = {2025},
  url          = {https://stackoverflow.com},
  note         = {Accessed: 2025-05-23}
}

@article{sqlstorm,
  title={SQLStorm: Taking Database Benchmarking into the LLM Era},
  author={Schmidt, Tobias and Leis, Viktor and Boncz, Peter and Neumann, Thomas},
  journal={Proceedings of the VLDB Endowment},
  volume={18},
  number={11},
  pages={4144--4157},
  year={2025},
  publisher={VLDB Endowment}
}

@article{slabcity,
  title={SlabCity: Whole-Query Optimization using Program Synthesis},
  author={Dong, Rui and Liu, Jie and Zhu, Yuxuan and Yan, Cong and Mozafari, Barzan and Wang, Xinyu},
  journal={Proceedings of the VLDB Endowment},
  volume={16},
  number={11},
  pages={3151--3164},
  year={2023},
  publisher={VLDB Endowment}
}

@article{ding2021dsb,
  title={DSB: A decision support benchmark for workload-driven and traditional database systems},
  author={Ding, Bailu and Chaudhuri, Surajit and Gehrke, Johannes and Narasayya, Vivek},
  journal={Proceedings of the VLDB Endowment},
  volume={14},
  number={13},
  pages={3376--3388},
  year={2021},
  publisher={VLDB Endowment}
}

@article{lr,
  title={A learned query rewrite system using monte carlo tree search},
  author={Zhou, Xuanhe and Li, Guoliang and Chai, Chengliang and Feng, Jianhua},
  journal={Proceedings of the VLDB Endowment},
  volume={15},
  number={1},
  pages={46--58},
  year={2021},
  publisher={VLDB Endowment}
}

@article{cloud,
  title={Cloud-native database systems at Alibaba: Opportunities and challenges},
  author={Li, Feifei},
  journal={Proceedings of the VLDB Endowment},
  volume={12},
  number={12},
  pages={2263--2272},
  year={2019},
  publisher={VLDB Endowment}
}

@article{metagpt,
  title={Metagpt: Meta programming for multi-agent collaborative framework},
  author={Hong, Sirui and Zheng, Xiawu and Chen, Jonathan and Cheng, Yuheng and Wang, Jinlin and Zhang, Ceyao and Wang, Zili and Yau, Steven Ka Shing and Lin, Zijuan and Zhou, Liyang and others},
  journal={arXiv preprint arXiv:2308.00352},
  volume={3},
  number={4},
  pages={6},
  year={2023}
}

@article{cte1,
  title={Optimization of common table expressions in mpp database systems},
  author={El-Helw, Amr and Raghavan, Venkatesh and Soliman, Mohamed A and Caragea, George and Gu, Zhongxian and Petropoulos, Michalis},
  journal={Proceedings of the VLDB Endowment},
  volume={8},
  number={12},
  pages={1704--1715},
  year={2015},
  publisher={VLDB Endowment}
}

@inproceedings{cte2,
  title={Efficient exploitation of similar subexpressions for query processing},
  author={Zhou, Jingren and Larson, Per-Ake and Freytag, Johann-Christoph and Lehner, Wolfgang},
  booktitle={Proceedings of the 2007 ACM SIGMOD international conference on Management of data},
  pages={533--544},
  year={2007}
}

@article{lr_demo,
  author    = {Xuanhe Zhou and
               Guoliang Li and
               Jianming Wu and
               Jiesi Liu and
               Zhaoyan Sun and
               Xinning Zhang},
  title     = {A Learned Query Rewrite System},
  journal   = {Proc. {VLDB} Endow.},
  year      = {2023},
}

@article{bm25,
  title={The probabilistic relevance framework: BM25 and beyond},
  author={Robertson, Stephen and Zaragoza, Hugo and others},
  journal={Foundations and Trends{\textregistered} in Information Retrieval},
  volume={3},
  number={4},
  pages={333--389},
  year={2009},
  publisher={Now Publishers, Inc.}
}

@article{sentence-transformer,
  title={Sentence-bert: Sentence embeddings using siamese bert-networks},
  author={Reimers, Nils and Gurevych, Iryna},
  journal={arXiv preprint arXiv:1908.10084},
  year={2019}
}

@article{LLMSQLSolverCL,
  title={LLM-SQL-Solver: Can LLMs Determine SQL Equivalence?},
  author={Fuheng Zhao and Lawrence Lim and Ishtiyaque Ahmad and Divyakant Agrawal and A. El Abbadi},
  journal={ArXiv},
  year={2023},
  volume={abs/2312.10321},
  url={https://api.semanticscholar.org/CorpusID:266348267}
}

@inproceedings{leo,
  title={LEO-DB2's learning optimizer},
  author={Stillger, Michael and Lohman, Guy M and Markl, Volker and Kandil, Mokhtar},
  booktitle={VLDB},
  volume={1},
  pages={19--28},
  year={2001}
}

@article{llm-r2,
  title={LLM-R2: A Large Language Model Enhanced Rule-based Rewrite System for Boosting Query Efficiency},
  author={Li, Zhaodonghui and Yuan, Haitao and Wang, Huiming and Cong, Gao and Bing, Lidong},
  journal={arXiv preprint arXiv:2404.12872},
  year={2024}
}

@inproceedings{apache,
  title={Apache calcite: A foundational framework for optimized query processing over heterogeneous data sources},
  author={Begoli, Edmon and Camacho-Rodr{\'\i}guez, Jes{\'u}s and Hyde, Julian and Mior, Michael J and Lemire, Daniel},
  booktitle={Proceedings of the 2018 International Conference on Management of Data},
  pages={221--230},
  year={2018}
}

@article{graefe1993query,
  title={Query evaluation techniques for large databases},
  author={Graefe, Goetz},
  journal={ACM Computing Surveys (CSUR)},
  volume={25},
  number={2},
  pages={73--169},
  year={1993},
  publisher={ACM New York, NY, USA}
}

@article{d-bot,
  title={D-bot: Database diagnosis system using large language models},
  author={Zhou, Xuanhe and Li, Guoliang and Sun, Zhaoyan and Liu, Zhiyuan and Chen, Weize and Wu, Jianming and Liu, Jiesi and Feng, Ruohang and Zeng, Guoyang},
  journal={arXiv preprint arXiv:2312.01454},
  year={2023}
}

@article{liu2023reason,
  title={Reason for future, act for now: A principled framework for autonomous llm agents with provable sample efficiency},
  author={Liu, Zhihan and Hu, Hao and Zhang, Shenao and Guo, Hongyi and Ke, Shuqi and Liu, Boyi and Wang, Zhaoran},
  journal={arXiv preprint arXiv:2309.17382},
  year={2023}
}

@misc{deepseek_doc,
  author = {{DeepSeek-AI}},
  title = {{The Temperature Parameter
}},
  year = {2025},
  url = {https://api-docs.deepseek.com/quick_start/parameter_settings},
  note = {Accessed: 2025-03-14}
}

@article{kuratov2024babilong,
  title={Babilong: Testing the limits of llms with long context reasoning-in-a-haystack},
  author={Kuratov, Yury and Bulatov, Aydar and Anokhin, Petr and Rodkin, Ivan and Sorokin, Dmitry and Sorokin, Artyom and Burtsev, Mikhail},
  journal={Advances in Neural Information Processing Systems},
  volume={37},
  pages={106519--106554},
  year={2024}
}

@article{jiang2023longllmlingua,
  title={Longllmlingua: Accelerating and enhancing llms in long context scenarios via prompt compression},
  author={Jiang, Huiqiang and Wu, Qianhui and Luo, Xufang and Li, Dongsheng and Lin, Chin-Yew and Yang, Yuqing and Qiu, Lili},
  journal={arXiv preprint arXiv:2310.06839},
  year={2023}
}

@inproceedings{shi2023large,
  title={Large language models can be easily distracted by irrelevant context},
  author={Shi, Freda and Chen, Xinyun and Misra, Kanishka and Scales, Nathan and Dohan, David and Chi, Ed H and Sch{\"a}rli, Nathanael and Zhou, Denny},
  booktitle={International Conference on Machine Learning},
  pages={31210--31227},
  year={2023},
  organization={PMLR}
}

@misc{anthropic_doc,
  author = {{Anthropic}},
  title = {{Create a Text Completion
}},
  year = {2025},
  url = {https://docs.anthropic.com/en/api/complete},
  note = {Accessed: 2025-03-14}
}

@article{singh2024exploring,
  title={Exploring the Use of LLMs for SQL Equivalence Checking},
  author={Singh, Rajat and Bedathur, Srikanta},
  journal={arXiv preprint arXiv:2412.05561},
  year={2024}
}

@article{mdp,
  title={Markov decision processes},
  author={Puterman, Martin L},
  journal={Handbooks in operations research and management science},
  volume={2},
  pages={331--434},
  year={1990},
  publisher={Elsevier}
}

@article{tutorial,
  title={A tutorial on llm reasoning: Relevant methods behind chatgpt o1},
  author={Wang, Jun},
  journal={arXiv preprint arXiv:2502.10867},
  year={2025}
}

@article{querybooster,
  title={QueryBooster: Improving SQL Performance Using Middleware Services for Human-Centered Query Rewriting},
  author={Bai, Qiushi and Alsudais, Sadeem and Li, Chen},
  journal={arXiv preprint arXiv:2305.08272},
  year={2023}
}

@article{genrewrite,
  title={Query Rewriting via Large Language Models},
  author={Liu, Jie and Mozafari, Barzan},
  journal={arXiv preprint arXiv:2403.09060},
  year={2024}
}

@article{wetune,
author = {Wang, Zhaoguo and Zhou, Zhou and Yang, Yicun and Ding, Haoran and Hu, Gansen and Ding, Ding and Tang, Chuzhe and Chen, Haibo and Li, Jinyang},
title = {WeTune: Automatic Discovery and Verification of Query Rewrite Rules},
year = {2022},
isbn = {9781450392495},
publisher = {Association for Computing Machinery},
address = {New York, NY, USA},
url = {https://doi.org/10.1145/3514221.3526125},
doi = {10.1145/3514221.3526125},
booktitle = {Proceedings of the 2022 International Conference on Management of Data},
pages = {94–107},
numpages = {14},
location = {Philadelphia, PA, USA},
series = {SIGMOD '22}
}

@article{multi-agent-chanllenges,
  title={LLM multi-agent systems: Challenges and open problems},
  author={Han, Shanshan and Zhang, Qifan and Yao, Yuhang and Jin, Weizhao and Xu, Zhaozhuo and He, Chaoyang},
  journal={arXiv preprint arXiv:2402.03578},
  year={2024}
}

@INPROCEEDINGS{131472,
  author={Finance, B. and Gardarin, G.},
  booktitle={[1991] Proceedings. Seventh International Conference on Data Engineering}, 
  title={A rule-based query rewriter in an extensible DBMS}, 
  year={1991},
  volume={},
  number={},
  pages={248-256},
  doi={10.1109/ICDE.1991.131472}}

@misc{wu2025agenticreasoningreasoningllms,
      title={Agentic Reasoning: Reasoning LLMs with Tools for the Deep Research}, 
      author={Junde Wu and Jiayuan Zhu and Yuyuan Liu},
      year={2025},
      eprint={2502.04644},
      archivePrefix={arXiv},
      primaryClass={cs.AI},
      url={https://arxiv.org/abs/2502.04644}, 
}

@article{10.1145/3485496,
author = {Nandi, Chandrakana and Willsey, Max and Zhu, Amy and Wang, Yisu Remy and Saiki, Brett and Anderson, Adam and Schulz, Adriana and Grossman, Dan and Tatlock, Zachary},
title = {Rewrite rule inference using equality saturation},
year = {2021},
issue_date = {October 2021},
publisher = {Association for Computing Machinery},
address = {New York, NY, USA},
volume = {5},
number = {OOPSLA},
url = {https://doi.org/10.1145/3485496},
doi = {10.1145/3485496},
journal = {Proc. ACM Program. Lang.},
month = oct,
articleno = {119},
numpages = {28}
}

@misc{postgresqldocs,
  title = {PostgreSQL: Documentation},
  howpublished = {\url{https://www.postgresql.org/docs/}},
  note = {Accessed: 2025-03-15}
}

@misc{postgresqlcte,
  title = {PostgreSQL: Common Table Expression Materialization},
  howpublished = {\url{https://www.postgresql.org/docs/14/queries-with.html}},
  note = {Accessed: 2025-03-17}
}

@inproceedings{bai-etal-2024-longbench,
    title = "{L}ong{B}ench: A Bilingual, Multitask Benchmark for Long Context Understanding",
    author = "Bai, Yushi  and
      Lv, Xin  and
      Zhang, Jiajie  and
      Lyu, Hongchang  and
      Tang, Jiankai  and
      Huang, Zhidian  and
      Du, Zhengxiao  and
      Liu, Xiao  and
      Zeng, Aohan  and
      Hou, Lei  and
      Dong, Yuxiao  and
      Tang, Jie  and
      Li, Juanzi",
    editor = "Ku, Lun-Wei  and
      Martins, Andre  and
      Srikumar, Vivek",
    booktitle = "Proceedings of the 62nd Annual Meeting of the Association for Computational Linguistics (Volume 1: Long Papers)",
    month = aug,
    year = "2024",
    address = "Bangkok, Thailand",
    publisher = "Association for Computational Linguistics",
    url = "https://aclanthology.org/2024.acl-long.172/",
    doi = "10.18653/v1/2024.acl-long.172",
    pages = "3119--3137"
}

@inproceedings{tpcds,
  title={The Making of TPC-DS.},
  author={Nambiar, Raghunath Othayoth and Poess, Meikel},
  booktitle={VLDB},
  volume={6},
  pages={1049--1058},
  year={2006}
}

@misc{mysql_docs,
  title = {MySQL Documentation},
  howpublished = {\url{https://dev.mysql.com/doc/}},
  note = {Accessed: 2025-03-15}
}

@misc{sqlserver_docs,
  title = {SQL Server technical documentation - SQL Server},
  howpublished = {\url{https://learn.microsoft.com/en-us/sql/sql-server/}},
  note = {Accessed: 2025-03-15}
}

@misc{oracle_db_docs,
  title = {Oracle Database Documentation - Oracle Database},
  howpublished = {\url{https://docs.oracle.com/en/database/oracle/oracle-database}},
  note = {Accessed: 2025-03-15}
}

@article{liu2023your,
  title={Is your code generated by chatgpt really correct? rigorous evaluation of large language models for code generation},
  author={Liu, Jiawei and Xia, Chunqiu Steven and Wang, Yuyao and Zhang, Lingming},
  journal={Advances in Neural Information Processing Systems},
  volume={36},
  pages={21558--21572},
  year={2023}
}

@article{mdagent,
  title={Mdagents: An adaptive collaboration of llms for medical decision-making},
  author={Kim, Yubin and Park, Chanwoo and Jeong, Hyewon and Chan, Yik S and Xu, Xuhai and McDuff, Daniel and Lee, Hyeonhoon and Ghassemi, Marzyeh and Breazeal, Cynthia and Park, Hae W},
  journal={Advances in Neural Information Processing Systems},
  volume={37},
  pages={79410--79452},
  year={2024}
}

@article{thompson2023large,
  title={Large language models with retrieval-augmented generation for zero-shot disease phenotyping},
  author={Thompson, Will E and Vidmar, David M and De Freitas, Jessica K and Pfeifer, John M and Fornwalt, Brandon K and Chen, Ruijun and Altay, Gabriel and Manghnani, Kabir and Nelsen, Andrew C and Morland, Kellie and others},
  journal={arXiv preprint arXiv:2312.06457},
  year={2023}
}

@article{multi1,
  title={Multi-agent collaboration: Harnessing the power of intelligent llm agents},
  author={Talebirad, Yashar and Nadiri, Amirhossein},
  journal={arXiv preprint arXiv:2306.03314},
  year={2023}
}

@article{multi2,
  title={Reflective multi-agent collaboration based on large language models},
  author={Bo, Xiaohe and Zhang, Zeyu and Dai, Quanyu and Feng, Xueyang and Wang, Lei and Li, Rui and Chen, Xu and Wen, Ji-Rong},
  journal={Advances in Neural Information Processing Systems},
  volume={37},
  pages={138595--138631},
  year={2024}
}

@article{multi3,
  title={Camel: Communicative agents for" mind" exploration of large language model society},
  author={Li, Guohao and Hammoud, Hasan and Itani, Hani and Khizbullin, Dmitrii and Ghanem, Bernard},
  journal={Advances in Neural Information Processing Systems},
  volume={36},
  pages={51991--52008},
  year={2023}
}

@article{multi4,
  title={Communicative agents for software development},
  author={Qian, Chen and Cong, Xin and Yang, Cheng and Chen, Weize and Su, Yusheng and Xu, Juyuan and Liu, Zhiyuan and Sun, Maosong},
  journal={arXiv preprint arXiv:2307.07924},
  volume={6},
  number={3},
  pages={1},
  year={2023}
}

@article{asai2024self,
  title={Self-rag: Learning to retrieve, generate, and critique through self-reflection},
  author={Asai, Akari and Wu, Zeqiu and Wang, Yizhong and Sil, Avirup and Hajishirzi, Hannaneh},
  year={2024},
  publisher={ICLR}
}

@misc{yang2024sweagentagentcomputerinterfacesenable,
      title={SWE-agent: Agent-Computer Interfaces Enable Automated Software Engineering}, 
      author={John Yang and Carlos E. Jimenez and Alexander Wettig and Kilian Lieret and Shunyu Yao and Karthik Narasimhan and Ofir Press},
      year={2024},
      eprint={2405.15793},
      archivePrefix={arXiv},
      primaryClass={cs.SE},
      url={https://arxiv.org/abs/2405.15793}, 
}

@article{r-bot,
  title={R-Bot: An LLM-based Query Rewrite System},
  author={Sun, Zhaoyan and Zhou, Xuanhe and Li, Guoliang},
  journal={Proceedings of the VLDB Endowment},
  volume={18},
  number={12},
  pages={ 5031 -- 5044},
  year={2025},
}

@inproceedings{volcano,
author = {Graefe, Goetz and McKenna, William J.},
title = {The Volcano Optimizer Generator: Extensibility and Efficient Search},
year = {1993},
isbn = {0818635703},
publisher = {IEEE Computer Society},
address = {USA},
booktitle = {Proceedings of the Ninth International Conference on Data Engineering},
pages = {209–218},
numpages = {10}
}

@misc{pg,
  author       = {{PostgreSQL Global Development Group}},
  title        = {{PostgreSQL: The world's most advanced open source database}},
  year         = {2024},
  howpublished = {\url{https://www.postgresql.org/}},
  note         = {Retrieved April 11, 2025}
}

@article{autosteer,
  title={Autosteer: Learned query optimization for any sql database},
  author={Anneser, Christoph and Tatbul, Nesime and Cohen, David and Xu, Zhenggang and Pandian, Prithviraj and Laptev, Nikolay and Marcus, Ryan},
  journal={Proceedings of the VLDB Endowment},
  volume={16},
  number={12},
  pages={3515--3527},
  year={2023},
  publisher={VLDB Endowment}
}

@article{fastgres,
  title={Fastgres: Making learned query optimizer hinting effective},
  author={Woltmann, Lucas and Thiessat, Jerome and Hartmann, Claudio and Habich, Dirk and Lehner, Wolfgang},
  journal={Proceedings of the VLDB Endowment},
  volume={16},
  number={11},
  pages={3310--3322},
  year={2023},
  publisher={VLDB Endowment}
}

@article{gptuner,
author = {Lao, Jiale and Wang, Yibo and Li, Yufei and Wang, Jianping and Zhang, Yunjia and Cheng, Zhiyuan and Chen, Wanghu and Tang, Mingjie and Wang, Jianguo},
title = {GPTuner: A Manual-Reading Database Tuning System via GPT-Guided Bayesian Optimization},
year = {2024},
issue_date = {April 2024},
publisher = {VLDB Endowment},
volume = {17},
number = {8},
issn = {2150-8097},
url = {https://doi.org/10.14778/3659437.3659449},
doi = {10.14778/3659437.3659449},
journal = {Proc. VLDB Endow.},
month = apr,
pages = {1939–1952},
numpages = {14}
}

@inproceedings{gptuner-demo,
  title={A demonstration of gptuner: A gpt-based manual-reading database tuning system},
  author={Lao, Jiale and Wang, Yibo and Li, Yufei and Wang, Jianping and Zhang, Yunjia and Cheng, Zhiyuan and Chen, Wanghu and Zhou, Yuanchun and Tang, Mingjie and Wang, Jianguo},
  booktitle={Companion of the 2024 International Conference on Management of Data},
  pages={504--507},
  year={2024}
}

@article{gptuner-record,
  title={GPTuner: An LLM-Based Database Tuning System},
  author={Lao, Jiale and Wang, Yibo and Li, Yufei and Wang, Jianping and Zhang, Yunjia and Cheng, Zhiyuan and Chen, Wanghu and Tang, Mingjie and Wang, Jianguo},
  journal={ACM SIGMOD Record},
  volume={54},
  number={1},
  pages={101--110},
  year={2025},
  publisher={ACM New York, NY, USA}
}

@article{llmidx,
  title={LLMIdxAdvis: Resource-Efficient Index Advisor Utilizing Large Language Model},
  author={Zhao, Xinxin and Li, Haoyang and Zhang, Jing and Huang, Xinmei and Zhang, Tieying and Chen, Jianjun and Shi, Rui and Li, Cuiping and Chen, Hong},
  journal={arXiv preprint arXiv:2503.07884},
  year={2025}
}

@article{llmqo,
  title={Can Large Language Models Be Query Optimizer for Relational Databases?},
  author={Tan, Jie and Zhao, Kangfei and Li, Rui and Yu, Jeffrey Xu and Piao, Chengzhi and Cheng, Hong and Meng, Helen and Zhao, Deli and Rong, Yu},
  journal={arXiv preprint arXiv:2502.05562},
  year={2025}
}

@article{text2sql2,
  title={Text-to-sql empowered by large language models: A benchmark evaluation},
  author={Gao, Dawei and Wang, Haibin and Li, Yaliang and Sun, Xiuyu and Qian, Yichen and Ding, Bolin and Zhou, Jingren},
  journal={arXiv preprint arXiv:2308.15363},
  year={2023}
}

@article{eq_np,
author = {Aho, A. V. and Sagiv, Y. and Ullman, J. D.},
title = {Equivalences among Relational Expressions},
journal = {SIAM Journal on Computing},
volume = {8},
number = {2},
pages = {218-246},
year = {1979},
doi = {10.1137/0208017},
URL = { 
        https://doi.org/10.1137/0208017
},
eprint = { 
        https://doi.org/10.1137/0208017
}
}

@inproceedings{steeringqo,
  title={Steering query optimizers: A practical take on big data workloads},
  author={Negi, Parimarjan and Interlandi, Matteo and Marcus, Ryan and Alizadeh, Mohammad and Kraska, Tim and Friedman, Marc and Jindal, Alekh},
  booktitle={Proceedings of the 2021 International Conference on Management of Data},
  pages={2557--2569},
  year={2021}
}

@inproceedings{SQLBarber,
author = {Lao, Jiale and Trummer, Immanuel},
title = {Demonstrating SQLBarber: Leveraging Large Language Models to Generate Customized and Realistic SQL Workloads},
year = {2025},
isbn = {9798400715648},
publisher = {Association for Computing Machinery},
address = {New York, NY, USA},
url = {https://doi.org/10.1145/3722212.3725101},
doi = {10.1145/3722212.3725101},
booktitle = {Companion of the 2025 International Conference on Management of Data},
pages = {151–154},
numpages = {4},
location = {Berlin, Germany},
series = {SIGMOD/PODS '25}
}

@misc{lao2025sqlbarberleveraginglargelanguage,
      title={SQLBarber: A System Leveraging Large Language Models to Generate Customized and Realistic SQL Workloads}, 
      author={Jiale Lao and Immanuel Trummer},
      year={2025},
      eprint={2507.06192},
      archivePrefix={arXiv},
      primaryClass={cs.DB},
      url={https://arxiv.org/abs/2507.06192}, 
}

@article{zhang2023unified,
  title={A unified and efficient coordinating framework for autonomous DBMS tuning},
  author={Zhang, Xinyi and Chang, Zhuo and Wu, Hong and Li, Yang and Chen, Jia and Tan, Jian and Li, Feifei and Cui, Bin},
  journal={Proceedings of the ACM on Management of Data},
  volume={1},
  number={2},
  pages={1--26},
  year={2023},
  publisher={ACM New York, NY, USA}
}

@article{zhang2024holon,
  title={The Holon Approach for Simultaneously Tuning Multiple Components in a Self-Driving Database Management System with Machine Learning via Synthesized Proto-Actions},
  author={Zhang, William and Lim, Wan Shen and Butrovich, Matthew and Pavlo, Andrew},
  journal={Proceedings of the VLDB Endowment},
  volume={17},
  number={11},
  pages={3373--3387},
  year={2024},
  publisher={VLDB Endowment}
}

\end{document}